\documentclass[aps,twocolumn,pra,groupedaddress,superscriptaddress,nofootinbib,amsmath]{revtex4-2}

\usepackage{graphicx,amsmath,amssymb,float}
\usepackage{txfonts}
\usepackage{hyperref,bbm,times}
\usepackage[T1]{fontenc}
\usepackage{soul}
\usepackage{dsfont}
\usepackage{braket}
\usepackage[table]{xcolor}
\usepackage{epsfig}
\usepackage{color}
\usepackage{graphicx}
\usepackage{dcolumn}
\usepackage{bm}
\usepackage{commath}

\NewDocumentEnvironment{alignb}{b}{%
  \begin{align*}
  \refstepcounter{equation} #1 \tag{\theequation}
  \end{align*}
}{\ignorespacesafterend}

\allowdisplaybreaks

\usepackage{tikz}
\usetikzlibrary{
  arrows.meta,
  decorations.pathmorphing,
  positioning,
  shapes.symbols
}

\definecolor{tunnelred}{RGB}{220,50,47}
\definecolor{bathgreen}{RGB}{38,139,108}

\newcommand{\ld}{\hat{L}_{\sigma}^{\dagger}}
\newcommand{\lsig}{\hat{L}_{\sigma}}
\newcommand{\rd}{\hat{R}_{\sigma}^{\dagger}}
\newcommand{\rsig}{\hat{R}_{\sigma}}

\newcommand{\lupc}{\hat{L}_{\uparrow}^{\dagger}}
\newcommand{\rupc}{\hat{R}_{\uparrow}^{\dagger}}
\newcommand{\ldownc}{\hat{L}_{\downarrow}^{\dagger}}
\newcommand{\rdownc}{\hat{R}_{\downarrow}^{\dagger}}

\newcommand{\lupd}{\hat{L}_{\uparrow}}
\newcommand{\rupd}{\hat{R}_{\uparrow}}
\newcommand{\ldownd}{\hat{L}_{\downarrow}}
\newcommand{\rdownd}{\hat{R}_{\downarrow}}

\newcommand{\szr}{\hat{\sigma}_{z, R}}
\newcommand{\szl}{\hat{\sigma}_{z, L}}

\newcommand{\er}{\hat{\mathcal{E}}_R}
\newcommand{\el}{\hat{\mathcal{E}}_L}

\newcommand{\Vop}{\hat{V}(t)}

\newcommand{\al}{\hat{\alpha}_L}
\newcommand{\ar}{\hat{\alpha}_R}
\newcommand{\A}{\hat{F}_0}
\newcommand{\B}{\hat{F}_1}
\newcommand{\C}{\hat{F}_2}

\providecommand{\openone}{\leavevmode\hbox{\small1\kern-3.8pt\normalsize1}}
\usepackage{soul}
\graphicspath{{figure/}}
\newread\tmp

\newif\ifannote
\annotetrue 

\ifannote
    
    \newcommand{\anncomment}[3]{{\color{#1}[#2: #3]}}
\else
    
    \newcommand{\anncomment}[3]{}
\fi

\begin{document}

\title{Tunneling of bosonic qubits under local dephasing through microscopic approach}

\author{Alberto Ferrara}
\email{alberto.ferrara@unipa.it}
\affiliation{Dipartimento di Ingegneria, Universit\`{a} degli Studi di Palermo, Viale delle Scienze, 90128 Palermo, Italy}

\author{Farzam Nosrati}
\affiliation{Dipartimento di Ingegneria, Universit\`{a} degli Studi di Palermo, Viale delle Scienze, 90128 Palermo, Italy}
\affiliation{IMDEA Networks Institute, Madrid, Spain}

\author{Andrea Smirne}
\affiliation{Universit\`{a} degli Studi di Milano, Dipartimento di Fisica, Via Celoria 16, I-20133 Milano, Italy}
\affiliation{Istituto Nazionale di Fisica Nucleare, Sezione di Milano, via Celoria 16, 20133 Milan, Italy}

\author{Jyrki Piilo}
\affiliation{Department of Physics and Astronomy, University of Turku, FI-20014, Turun Yliopisto, Finland}

\author{Rosario Lo Franco}%
\affiliation{Dipartimento di Ingegneria, Universit\`{a} degli Studi di Palermo, Viale delle Scienze, 90128 Palermo, Italy}

\begin{abstract}
We present a microscopic derivation of a master equation for two-component bosons (bosonic qubits) which tunnel between spatially separated modes under local dephasing noise. Starting from the full system-bath Hamiltonian with Lorentzian coupling distributions, we analytically obtain a time-local master equation whose structure reveals intrinsic non-Markovian features and recovers the standard phenomenological dephasing model in the short-time limit. Comparison with exact pseudomode simulations confirms its validity beyond weak-coupling and Markovian regimes. We identify a resonance condition between tunneling and bath frequencies for which dephasing drives the system towards correlated steady states, stabilizing coherence and entanglement instead of suppressing them. These results establish a rigorous microscopic foundation for dephasing models in bosonic tunneling systems and reveal a noise-induced mechanism for steady-state entanglement.
\end{abstract}

\maketitle
\section{Introduction}
Indistinguishability of identical particles lies at the heart of quantum interference phenomena, enabling effects such as bosonic bunching and fermionic anti-bunching \cite{hong1987measurement,pauli2012general,zeilingerRMP2022}. Such interference relies on coherent tunneling between spatially separated mode, a process that can be described as a \emph{spatial deformation} operation, generating the wave function overlap required for indistinguishability-driven correlations. These correlations are harnessed through spatial deformation followed by post-selection via spatially localized measurements and classical communication, forming a protocol for the generation of entangled states \cite{lo2018indistinguishability,nosrati2024indistinguishability,mahdavipour2024generation,Blasiak2019}. Such entanglement has been demonstrated as a genuine quantum resource in experimental settings \cite{sun2020experimental,Barros2020OEntangling,Lee2022,WangPRA2022}, including quantum teleportation in photonic platforms \cite{sun2020experimental} and quantum metrology \cite{sun2022activation}. The same approach has been shown to protect or even distill entanglement from noisy initial \cite{nosrati2020robust,PiccoliniAdv2023,nosrati2024indistinguishability,piccolini2024robust,piccolini2024asymptotically}. Previous studies assumed perfect, noiseless spatial deformation but, like any quantum process, it is inevitably subject to decoherence \cite{breuer2002theory,rivas2012open,vacchini2024open}. Thus, understanding how noise affects the system during spatial deformation is essential to assess the robustness of quantum information processing based on coherent tunneling.

To describe the interplay between coherent tunneling and simultaneous local noise, the dynamics can be studied within an open quantum system framework, where decoherence and dissipation continuously affect the coherent evolution \cite{breuer2002theory, rivas2012open, vacchini2024open}. A fundamental tool for achieving such an effective description is provided by master equations \cite{gorini1976completely,lindblad1976generators}. In this context, phenomenological master equations are widely used. However, a microscopic derivation of the master equation is essential to rigorously connect system-environment interactions with the resulting effective dynamics and to assess the validity of commonly adopted approximations. 

In this work, we derive a microscopic, time-local master equation for any generic initial state of bosonic qubits tunneling between spatially separated modes while interacting with local bosonic dephasing baths characterized by Lorentzian spectral densities. This framework enables us to investigate how decoherence, which occurs during the tunneling process, affects paradigmatic indistinguishability-driven phenomena, including Hong-Ou-Mandel (HOM) interference and indistinguishability-based entanglement distillation. The system we study is equivalent to a pseudospin-1/2 Bose-Hubbard (BH) model with negligible on-site interaction \cite{zvonarev2009dynamical, yang2003rigorous, essler2010threshold,cazalilla2011one, stamper2013spinor}. In this widely studied scenario, effects of noise have been investigated and white-noise local dephasing is expected to suppress coherence, driving the system towards classical mixtures in the long-time limit. This prediction, usually based on phenomenological master equations, suggests that no steady-state quantum correlations can survive \cite{fazio2024many, vatre2023dynamics, poletti2012interaction, bouganne2018probing}. Our microscopic approach reveals an unreported dynamical regime exhibiting steady-state entanglement when the bath natural frequency is resonant with the tunneling amplitude, while recovering the expected loss of coherence outside of this resonance. Overall, these results provide a rigorous microscopic foundation for dephasing models in bosonic tunneling systems, uncovering intrinsic non-Markovian dynamics beyond the weak-coupling regime and identifying noise-induced steady entangled states.

The manuscript is organized as follows. 
In Sec.~{\ref{sec: Model}} we present the system, defining the formal framework and the physical processes we are going to study. In Sec.~{\ref{sec: results}} we outline the derivation of the master equation, describing some of its emerging features as well as the numerical methods adopted to validate it. After this, we show results for three specific settings. In Sec.~{\ref{sec: A}} we study HOM interference. Sec.~{\ref{sec: C}} illustrates the resonance mechanism through specific single-particle states. in Sec.~{\ref{sec: opposite}} we investigate the entanglement generation protocol and finally, we summarize our conclusions in Sec.~{\ref{sec: Concl}}.
\section{Model}
\label{sec: Model}
We focus on a system made of two distinct spatial regions populated by particles following bosonic statistics. Each particle is characterized by a spatial degree of freedom and an internal pseudospin degree of freedom, admitting two distinct values. The mathematical framework under which the following results are obtained is worked out in second quantization formalism. Moreover, we adopt the no-label approach \cite{LoFranco2016Quantum,compagnoRSA}. According to this, a state vector describing a single particle with pseudospin $\sigma$ localized in $X$ can be expressed as $\ket{\phi} = \ket{X\:\sigma}$, where $X = \{L, R\}$ labels two distinct spatial regions (left $L$, right $R$), and $\sigma = \{ \uparrow, \downarrow\}$ represents two pseudospin values. 
We define creation and annihilation operators $\hat{X}^{\dagger}_{\sigma}$ and $\hat{X}_{\sigma}$, acting on a generic state by creating or annihilating particles with a given pseudospin in a specific spatial region (e.g., $\hat{L}^{\dagger}_{\uparrow} \ket{0} = \ket{L\:{\uparrow}}$, where $\ket{0}$ is a vacuum state). Since we are considering bosons, these operators obey canonical commutation rules, $[\hat{X}_{\sigma}, \hat{X}^{\dagger}_{\sigma'}] = \delta_{\sigma, \sigma'}$,  $[\hat{X}_{\sigma}, \hat{X}_{\sigma'}]$ = 0, and $[\hat{X}^{\dagger}_{\sigma}, \hat{X}^{\dagger}_{\sigma'}]$ = 0. This means that bosonic ladder operators with different pseudospins commute, and the same applies for operators acting on different spatial regions.

\begin{figure}[t]
  \centering
  \begin{tikzpicture}[
    >=Stealth,
    font=\small,
    thick,
    scale=1.0,
    every node/.style={transform shape},
    bath/.style={
      cloud,
      draw=bathgreen,
      thick,
      cloud puffs=10,
      cloud ignores aspect,
      minimum width=2.2cm,
      minimum height=1.2cm,
      align=center,
      fill=bathgreen!15
    },
    particle/.style={
      circle,
      draw=black,
      very thick,
      fill=gray!30,
      minimum size=0.65cm,
      inner sep=0pt
    }
  ]
    \draw[tunnelred, very thick, fill=tunnelred!12]
      (-3.2,2.4) -- (-3.2,0.95)
      .. controls (-2.8,0.55) and (-1.6,0.55) .. (-1.2,0.95)
      -- (-1.2,2.4) -- cycle;
    \node[font=\large] at (-2.2,2.15) {$L$};

    \draw[tunnelred, very thick, fill=tunnelred!12]
      ( 1.2,2.4) -- ( 1.2,0.95)
      .. controls ( 1.6,0.55) and ( 2.8,0.55) .. ( 3.2,0.95)
      -- ( 3.2,2.4) -- cycle;
    \node[font=\large] at (2.2,2.15) {$R$};

    \node[particle] (particleL) at (-2.2,1.5) {};
    \draw[<->, very thick, line width=0.8pt] 
      ([yshift=-8pt]particleL.center) -- ([yshift=8pt]particleL.center);

    \node[particle] (particleR) at (2.2,1.5) {};
    \draw[<->, very thick, line width=0.8pt] 
      ([yshift=-8pt]particleR.center) -- ([yshift=8pt]particleR.center);

    \draw[<->, tunnelred, very thick, line width=1.2pt]
      (-0.45,1.8) -- (0.45,1.8)
      node[midway, above=3pt, font=\normalsize] {$J$}
      node[midway, black, above=-20pt, font=\normalsize] {\textbf{Tunneling}};

    \node[bath] (BL) at (-2.2,-1.0) {\small\textbf{Local}\\\small\textbf{Dephasing}};
    \node[bath] (BR) at ( 2.2,-1.0) {\small\textbf{Local}\\\small\textbf{Dephasing}};

    \coordinate (BL1) at (-2.8,-0.25);
    \coordinate (BL2) at (-1.55,-0.25);
    \coordinate (Target1) at (-2.80,0.65);
    \coordinate (Target2) at (-1.55,0.65);

    \draw[bathgreen, very thick, -Stealth, decorate,
          decoration={snake, amplitude=2pt, segment length=9pt, post length=2pt}]
      (BL1) -- (Target1);
      
    \draw[bathgreen, very thick, -Stealth, decorate,
          decoration={snake, amplitude=2pt, segment length=9pt, post length=2pt}]
      (BL2) -- (Target2);

    \coordinate (BR1) at (1.55,-0.25);
    \coordinate (BR2) at (2.80,-0.25);
    \coordinate (Target3) at (1.55,0.65);
    \coordinate (Target4) at (2.80,0.65);

    \draw[bathgreen, very thick, -Stealth, decorate,
          decoration={snake, amplitude=2pt, segment length=9pt, post length=2pt}]
      (BR1) -- (Target3);
      
    \draw[bathgreen, very thick, -Stealth, decorate,
          decoration={snake, amplitude=2pt, segment length=9pt, post length=2pt}]
      (BR2) -- (Target4);

  \end{tikzpicture}
  \caption{Illustration of the system. Two spatial regions $L$ and $R$ can exchange bosons via tunneling of amplitude $J$. Each region hosts a bosonic qubit (gray circle) with a given pseudospin (double-headed arrow) and is subject to local dephasing from its own bosonic bath.}
  \label{fig: setup}
\end{figure}

With $\hbar\equiv 1$, the free Hamiltonian of our system is $\hat{H}_S = \omega_s  \sum_{X = \{L, R\},\ \sigma = \{\uparrow, \downarrow\}} \hat{n}_{X\sigma} $, where $ \omega_s$ is the qubit transition frequency, which is the same for both particles, and $\hat{n}_{X\sigma} = \hat{X}_{\sigma}^{\dagger}\hat{X}_{\sigma}$ is the local number operator for a given pseudospin. 

The bosonic qubits in the two spatial regions interact through a tunneling process
\begin{equation}
    \hat{H}_D = \frac{J}{2} \sum_{\sigma = \{\uparrow, \downarrow\}}\left[ \ld \rsig +  \rd \lsig \right],
    \label{eq: deformation_hamiltonian}
\end{equation}
so that the tunneling can be controlled by both tunneling rate $J$ and evolution time $t$. This allows the particle to hop from one side to the other, leading to spatial deformation and overlap of their spatial wavefunctions, while leaving their pseudospin untouched. 

The qubits interact locally with two independent zero-temperature bosonic baths, respectively, as depicted in Fig.~\ref{fig: setup}. To model local pure-dephasing noise, we consider two distinct bosonic baths, whose overall free Hamiltonian is \(\hat{H}_{E} = \sum_{X = \{L, R \}}\int_0^{\infty} \omega \; \hat{b}_{X}^{\dagger}(\omega) \hat{b}_{X}(\omega) d\omega\), where $\hat{b}_{X}^{\dagger}(\omega)$ and $\hat{b}_{X}(\omega)$ are the bosonic creation and annihilation operators for mode frequency $\omega$ in the bath localized at site $X$. The system-bath interaction Hamiltonian is defined as
\begin{equation}
    \hat{H}_{SE} = \sum_{X = \{L, R \}} \left[ \hat{\sigma}_{z, X} \otimes \int_0^{\infty} g_{X} (\omega) [\hat{b}_{X}(\omega) + \hat{b}_{X}^{\dagger}(\omega)] d\omega \right] \text{,}
    \label{eq: system-bath_hamiltonian}
\end{equation}
where we have introduced a local dephasing operator $\hat{\sigma}_{z, X} = \hat{n}_{X\uparrow} - \hat{n}_{X\downarrow} = \hat{X}_{\uparrow}^{\dagger}\hat{X}_{\uparrow}-\hat{X}_{\downarrow}^{\dagger}\hat{X}_{\downarrow}$. This interaction linearly couples the number of excitations in one local region with its respective environment. The system-bath coupling $g_X(\omega)$ is described by a spectral Lorentzian distribution
\begin{equation}
    \abs{g_{X}(\omega)}^2 =\frac{g_0}{\pi}\frac{\lambda}{(\omega - \omega_0)^2 + \lambda^2},
\end{equation}
where $\lambda$ is the spectral width and $\omega_0$ is the central frequency of the bosonic bath mode. The parameter $g_0$ sets the overall coupling strength. This Hamiltonian contribution commutes with the free system Hamiltonian $\hat{H}_S$, but not with the deformation one $\hat{H}_{D}$, enabling nontrivial interplay between noise and tunneling dynamics. The total Hamiltonian can be written as $\hat{H}_\textrm{tot} = (\hat{H}_S + \hat{H}_D)\otimes \mathds{1}_E + \mathds{1}_S \otimes \hat{H}_{E} + \hat{H}_{SE}$. 

\section{Dynamical map}  
\label{sec: results}
In previous works, dephasing in bosonic systems, with or without tunneling, has been typically modeled via the following phenomenological master equation~\cite{fazio2024many,poletti2012interaction}:
\begin{equation}
    \dot{\hat{\varrho}}_S(t) = -i [\hat{H}_S + \hat{H}_D, \hat{\varrho}_S(t)] + \sum_{X = \{L, R\}} \Gamma_{X} \; D[\hat{\sigma}_{z,X}]\hat{\varrho}_S,
    \label{eq: mast_eq}
\end{equation}
where \(\hat{\varrho}_S(t)\) is the reduced density matrix of the system, $\Gamma_{X} \in \mathbb{R}^+$ are constant dephasing rates, and $D[\hat{A}]\hat{\varrho}=\hat{A}\hat{\varrho}\hat{A}^\dagger-\frac{1}{2}\{\hat{A}^\dagger\hat{A},\hat{\varrho}\}$ is the Lindblad dissipator. In absence of an internal (pseudospin) degree of freedom, the dissipative dynamics is described by a simple Markovian Lindblad dissipator determined by the local number operator for each spatial region. However, such phenomenological master equations lack a microscopical foundation and might not fully capture the underlying dynamics. To address this gap, we derive the master equation explicitly from a microscopic description of the system-environment interaction. 

The derivation begins by moving to the interaction picture using the standard unitary transformation \(\Vop=e^{i\hat{H_0}t}\hat{H}_{SE}e^{-i\hat{H_0}t}\) with $\hat{H}_0 = \hat{H}_S + \hat{H}_D + \hat{H}_E$. This yields the interaction-picture Hamiltonian
\begin{eqnarray}
    \Vop &=&\left(\hat{F}_{0}+\hat{F}_{1}\cos{\left(J t\right)}+\hat{F}_2\sin{\left(J t\right)}\right)\otimes\mathcal{E}_L(t) \nonumber \\
    &+&\left(\hat{F}_{0}-\hat{F}_{1}\cos{\left(J t\right)}-\hat{F}_2\sin{\left(J t\right)}\right)\otimes\mathcal{E}_R(t),
\end{eqnarray}
where
\(\hat{F}_{0}=\frac{1}{2}(\szl + \szr)\),  \(\hat{F}_{1}=\frac{1}{2}(\szl - \szr)\), $\hat{F}_2=\frac{i}{2}\left[\rupc \lupd - \lupc \rupd - \rdownc \ldownd + \ldownc \rdownd   \right]$, and the product $Jt$ can be viewed as a deformation angle. The time-dependent bath operator is $\mathcal{E}_X(t)= \int_0^{\infty} g_{X} (\omega) [\hat{b}_{X}(\omega) e^{-i \omega t} + \hat{b}_{X}^{\dagger} e^{+i \omega t}(\omega)] d\omega $. Following the standard procedure \cite{breuer2002theory}, we approximate the time derivative at second order in the coupling constant $g_0$, apply the Born approximation, and trace out the environmental degrees of freedom. Additionally, we assume that the initial state of the two local baths is the vacuum state $\ket{00_{\text{env}}} \coloneqq \ket{0}\otimes \ket{0} $. Globally, the master equation for the reduced density matrix $\hat{\varrho}^I_S(t)$, in the interaction picture, is
\begin{eqnarray}\label{ME1}
    \dot{\hat{\varrho}}_S^I(t) &=& -i [\hat{H}_\mathrm{can}(t), \hat{\varrho}_S^I(t) ]\nonumber\\ &+&\sum_{j, k = 0}^{2} \Gamma_{jk}(t) \left[ F_j \hat{\varrho}_S^I(t) F_k - \frac{1}{2} \{ \hat{\varrho}_S^I(t), F_k F_j\}\right]\text{,}
\end{eqnarray}
where $\hat{H}_{\mathrm{can}}(t)$ and $\Gamma_{jk}(t)$ denote the system Hamiltonian in canonical form and the damping matrix elements, respectively. We remark that this master equation is now expressed in a canonical form \cite{gorini1976completely}, which guarantees trace preservation and Hermiticity of the evolved density matrix \cite{manzano2020short}. Additionally, by studying the eigenvalues of the matrix $\Gamma(t)$ having elements $\Gamma_{jk}(t)$, one can investigate the physicality of the map \cite{bernal2021non} and characterize its non-Markovianity \cite{hall2014canonical}. A detailed step-by-step derivation of the master equation is provided in Appendix~\ref{sec: supp_mat_deriv}, along with the analytical expressions of $\hat{H}_\mathrm{can}(t)$ and the matrix elements $\Gamma_{ij}(t)$. 

\begin{figure}
    \centering
    \includegraphics[width=0.99\linewidth]{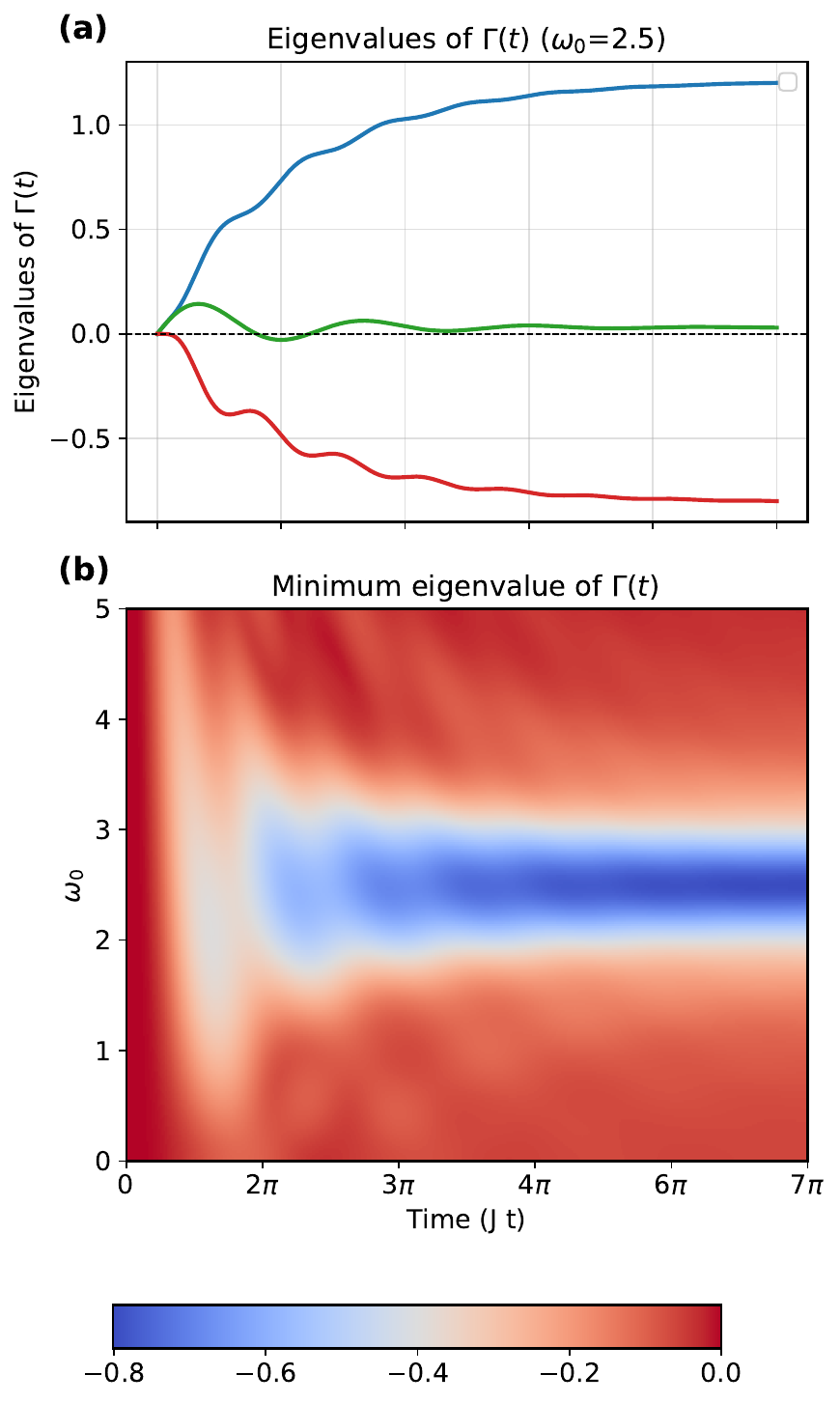}
    \caption{\textbf{(a)} Time evolution of the eigenvalues of the $3$x$3$ Kossakowski diffusion matrix $\Gamma$ at resonance ($\omega_0 = J$). \textbf{(b)} Minimum Eigenvalue analysis of $\Gamma$ over time, for varying values of the average value of the Lorentzian coupling distribution $\omega_0$. $\lambda = 0.5, g_0 = 0.1, J = 2.5$.}
    \label{fig: eigenvalues}
\end{figure}
The master equation of Eq.~(\ref{ME1}) can be numerically diagonalized, leading us to
\begin{equation}
    \dot{\hat{\varrho}} = -i [\hat{H}_{\text{can}}(t), \hat{\varrho} ] + \sum_{i \in \{ 0, -, +\} }\gamma_i(t) \left[ \hat{\mathcal{O}}_i(t) \hat{\varrho} \hat{\mathcal{O}}^{\dagger}_i(t) - \frac{1}{2} \{\hat{\mathcal{O}}^{\dagger}_i(t) \hat{\mathcal{O}}_i(t), \hat{\varrho}\}\right]\text{.}
    \label{eq: diag_canonical_form}
\end{equation}
The master equation derived above is time-local, with three time-dependent coefficients that can assume negative values. This indicates that the dynamics is generally non-Markovian, according to most commonly used non-Markovianity measures \cite{breuer2009measure, rivas2010entanglement, hall2014canonical}. Specifically, one of the three eigenvalues of $\Gamma(t)$, denoted as $\gamma_-(t)$, as shown in Fig.~\ref{fig: eigenvalues}, is always negative for the parameter range we studied, which is associated with a phenomenon known as eternal non-Markovianity \cite{megier2017eternal}.
The three jump operators can be obtained as $\hat{\mathcal{O}}_i =\sum_k U_{k}^i \hat{F}_k$, where $U_{k}^i$ are the elements of the diagonalization matrix $U$, with the index $i \in \{ 0, -, +\}$ corresponding to one of the three eigenvalues $\gamma_i(t)$ and the index $j \in \{0,1,2\}$ related to the old jump operators $\hat{F}_k$. We get $\hat{\mathcal{O}}_0 (t) = \A$, $\hat{\mathcal{O}}_+ (t) = U_1^+ (t) \B + U_2^+ (t) \C$ and $\hat{\mathcal{O}}_- (t) = U_1^- (t) \B + U_2^- (t) \C$. While the total number operator $\hat{F}_0$, associated with the eigenvalue $\gamma_0(t)$, decouples from the other two jump operators, we observe a competition between $\hat{\mathcal{O}}_+$ and $\hat{\mathcal{O}}_-$, since they have the same operatorial structure, while their respective eigenvalues $\gamma_+$ and $\gamma_-$ have opposite sign. Due to this, the master equation exhibits small periodic violation of another, non-equivalent indicator of non-Markovianity, namely P-divisibility, which expresses information backflow to the open system \cite{vacchini2011markovianity, breuer2016colloquium, benatti2024quantum}. The degree of non-Markovianity of the map can be controlled through the parameters of the spectral distribution, namely the coupling strength $g_0$ and the spectral width $\lambda$, as well as the central frequency $\omega_0$. More details can be found in Appendix~\ref{sec: non-mark}, where we show some explicit scenarios where P-divisibility is violated and we discuss in more detail the behavior of the diagonalization matrix $U$. 

In the following, we analyze one and two-particle states to compare the phenomenological and microscopic master equations, showing previously unreported dynamical regimes. The resulting dynamics are compared with numerical simulations of the full system--environment evolution. However, simulating a bipartite system coupled to a continuum of environmental frequencies is challenging, as it formally requires an infinite number of bosonic modes. To provide a reliable benchmark for the exact solution, we adopt the pseudomode method~\cite{garraway1997nonperturbative,dalton2001theory,mazzola2009pseudomodes, tamascelli2018nonperturbative, chen2019markovian,pleasance2020generalized,zhou2024systematic, menczel2024non}, where the original bath is replaced by a finite (possibly small) set of damped modes. A detailed description of its implementation is provided in the Appendix~\ref{sec: pseudo_mode}. 
We remark that the master equation, as well as the pseudomode description of the system-environment ensemble, have been derived without any specific constraints on the initial state. Thus they all admit, in principle, any number of particles over the two spatial regions, with initial states that can be either in any pure quantum superposition or mixed. An extra example of possible dynamics which explores multiple particle sectors of the Hilbert space is reported in Appendix~\ref{sec: example}.
We consider bosonic states with equal-pseudospin and opposite-pseudospin. Due to the specific system-environment correlation functions, two regimes emerge based on the coupling spectrum central frequency $\omega_0$: \textit{on-resonance} for $\omega_0 \approx J$ and \textit{off-resonance} when $|\omega_0 - J| > \lambda$. Following known dephasing models \cite{paladino20141}, we showcase the off-resonance case by setting $\omega_0 \approx 0$, so that the coupling distribution has negligible overlap with $J$. In the presence of steady states (on-resonance case), we quantify the agreement between the state $\varrho$ obtained via master equation and the reduced state $\sigma$ stemming from the pseudomode method by means of the trace distance $\mathcal{T}(\hat{{\varrho}}, \hat{\sigma}) = \frac{1}{2} \text{Tr}[\sqrt{(\hat{{\varrho}} - \hat{\sigma})^{\dagger} (\hat{{\varrho}} - \hat{\sigma}) }]$ \cite{nielsen2010quantum}.

\section{Equal-pseudospin bosons}
\label{sec: A}
We start by examining the HOM effect. In absence of noise, the pure separable initial state $\ket{L\,\sigma,R\,\sigma}$ evolves unitarily as 
\(\ket{\Psi(t)}=e^{i\hat{H}_Dt}\ket{L\:{\sigma}, R,{\sigma}}=\frac{i}{\sqrt{2}}\sin{(Jt)}\left(\ket{L\:\sigma,L\:\sigma}+\ket{R\:\sigma,R\:\sigma}\right)
+\cos{(Jt)}\ket{L\:\sigma,R\:\sigma}\). The noiseless dynamics generates a periodic oscillation between the initial separable state and a coherent NOON-like state. Since the dynamics leaves the pseudospin degree of freedom unaffected, this input state follows the same behavior as that of two generic spinless bosons under a continuous spatial deformation process. 
\begin{figure}[t!]
    \centering
    \includegraphics[width=0.99\linewidth]{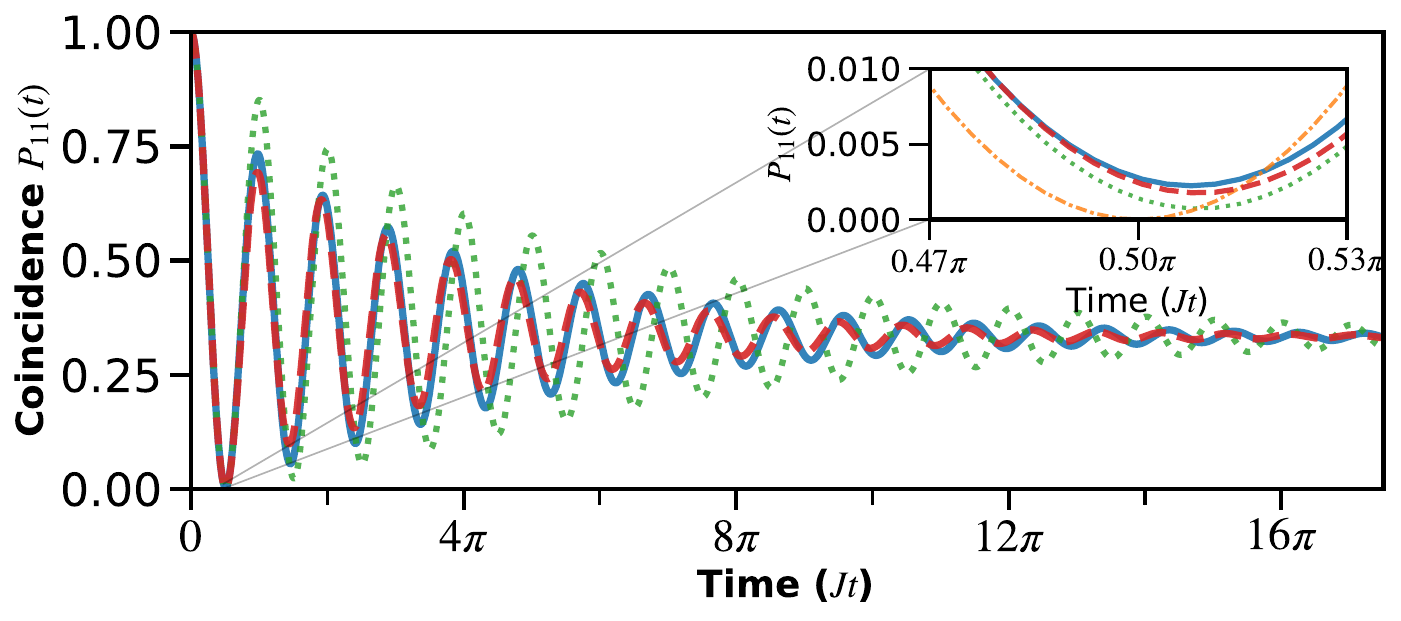}
    \caption{Off-resonance dynamics of coincidence probability for equal-pseudospin bosons, as a function of the deformation angle $Jt$, from numerical simulations using the full master equation (solid blue line), the phenomenological master equation (dotted green line), and the pseudomode evolution (dashed red line). The inset displays a zoom of the main plot around the first minimum of the coincidence probability, also showing the noiseless evolution (dash-dotted orange line). Parameters: $J = 2.5$, $g_0=0.5 $, $\lambda=1.8$, $\omega_0=0$, $\omega_s=1$, $\hbar = 1$.}
    \label{fig: OFF_spinless}
\end{figure}
To investigate the main features of the HOM effect, we track the coincidence probability, corresponding to the probability of detecting one boson per region, defined as
\(P_{11}(t)=\mathrm{Tr}\left(\hat{\varrho}_S\hat{\Pi}_{LR}\right)\), where $\hat{\Pi}_{LR} =  \ket{L\:{\sigma}, R,{\sigma}}\bra{L\:{\sigma}, R,{\sigma}}$ is the projective measurement operator that selects one boson in each region. When the value of this specific observable vanishes at $Jt=\pi/2$, the bosons live in a two-particle spatial superposition over the two regions. This is usually known as bosonic bunching \cite{hong1987measurement}. 

The time evolution of \(P_{11}(t)\) in presence of noise, in the off-resonance case, is shown in Fig.~\ref{fig: OFF_spinless}, where we compare results obtained from the phenomenological and microscopic master equations, with the pseudomode method used as a benchmark. Qualitatively, dephasing produces two distinct effects: (i) a suppression of bosonic bunching and (ii) a time delay of the HOM dip, meaning that, in the presence of noise, a larger deformation angle $J t$ is required to reach maximum bunching. Finally, at long times the density matrix converges to a steady state that is a classical mixture of the states $\ket{L\:\sigma,L\:\sigma}, \ket{L\:\sigma,R\:\sigma}, \ket{R\:\sigma,R\:\sigma}$, which is the subset of the total Hilbert space in which two bosons with the same pseudospin occupy the two spatial modes.

\begin{figure}[t!]
    \centering
    \includegraphics[width=0.99\linewidth]{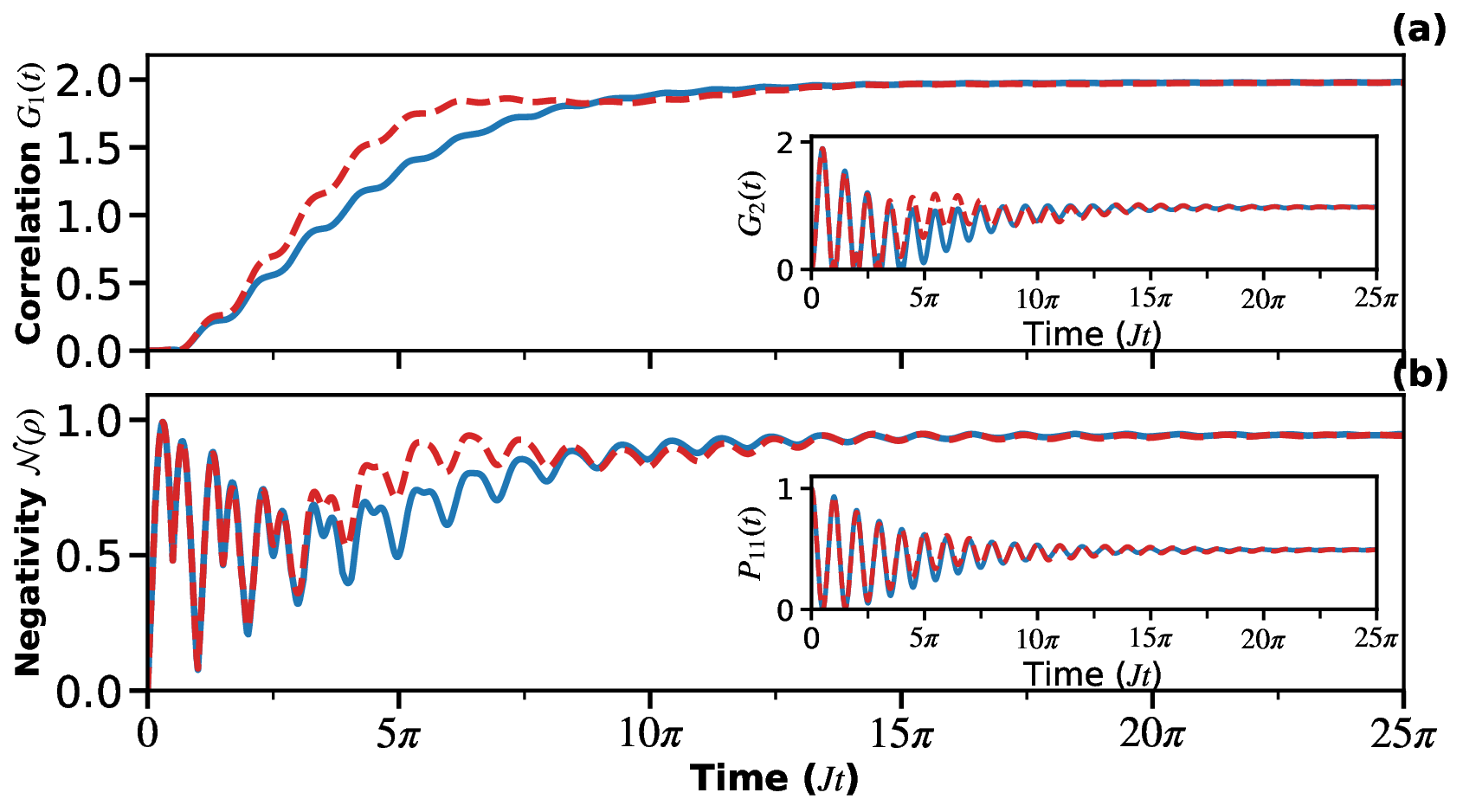}
    \caption{On-resonance evolution of first-order correlation function (a) and of entanglement negativity (b) for equal-pseudospin bosons from numerical simulations using the full master equation (solid blue line) and the pseudomode evolution (dashed red line). At the steady state, the trace distance is $\mathcal{T}= 0.010981$. Parameters: $g_0 = 0.1, \lambda=0.5, J= \omega_0 = 2.5, \omega_s = 1, \hbar = 1$.}
    \label{fig: ON_RES_spinless}
\end{figure}

The dynamics exhibits a different behavior at the steady state level in the on-resonance regime, when $J \approx \omega_0$. Specifically, tunneling between the two spatial regions is progressively inhibited, until the two-particle ensemble crystallizes into a correlated steady state. We quantify this trait through the first and second order correlation function $\mathcal{G}_n(t) = |\langle \lsig^n \hat{R}_{\sigma}^{\dagger n} + \text{h.c.}\rangle|(t)$ \cite{bach2004correlation}, plotted in Fig.~\ref{fig: ON_RES_spinless}~(a) and inset plots. Furthermore, we detect mode entanglement \cite{wiseman2003entanglement} between the two spatial degrees of freedom of the bosonic field, indicating that the occupations of the two modes are quantum correlated. This is quantified by the entanglement negativity, $\mathcal{N}(\hat{\varrho}) = \frac{|| \hat{\varrho}^{T_L} ||_1-1}{2}$, where $\hat{\varrho}^{T_L}$ is the partial transpose of the reduced density matrix with respect to either one of the two spatial modes \cite{vidal2002computable}. As shown in Fig.~\ref{fig: ON_RES_spinless}~(b), a significant amount of mode entanglement is generated and persists in the steady state. This behavior, emerging from the sole effect of the reduced dynamical evolution, lies beyond the reach of phenomenological master equations. In contrast, our microscopically derived master equation agrees closely with the pseudomode results.

The appearance of stationary entanglement under resonance condition can be interpreted within the pseudomode representation of the Lorentzian environment. In this picture, each structured reservoir is equivalent to a single damped harmonic mode of frequency $\omega_0$ and linewidth $\lambda$, which mediates coherent exchanges between the system and the residual Markovian bath. When the tunneling frequency matches the pseudomode frequency, the system and pseudomode become resonantly hybridized, giving rise to back and forth coherent phase oscillations. This resonance enables the environment to return phase information to the system on the same timescale as the intrinsic tunneling process, thereby suppressing the irreversible phase diffusion typical of dephasing noise. The finite pseudomode damping rate $\lambda$ ensures both a finite environmental memory time and a noise channel that stabilizes the coupled dynamics into a stationary regime. Hence, dephasing at resonance ceases to act as a purely decoherence process and induces a coherent feedback mechanism that drives the system toward a noise-assisted, entangled steady state.

\section{Resonance mechanism for single-particle dynamics}
\label{sec: C}
To shine light on the resonance behavior, we analyze the evolution of a single particle initially in a non-local superposition
\begin{equation}
    \ket{\Xi^{\pm}} = \frac{1}{\sqrt{2}}(\ket{L} \pm \ket{R}),
\end{equation}
where we have dropped the spin label $\sigma$, since the dynamics of a single-particle is not affected by the additional degree of freedom.
The peculiarity of these two initial states emerges directly from their invariance under the action of the system Hamiltonian $\hat{H}_S$.
More specifically, $\ket{\Xi^{\pm}}$ are the only eigenstates of the system Hamiltonian in the single-particle sector of the bosonic Hilbert space, with eigenenergies equal to $E_{\pm} = \omega_0 \pm \frac{J}{2}$. The energy gap between these two states is exactly $J$. When the system is closed (i.e. in the absence of noise), both states undergo no dynamical evolution. 
When turning on pure dephasing noise, due to the symmetry between the two spatial regions and due to the particle-number conserving dynamics, the diagonal elements of the density matrix associated to this state cannot change. Thus, the only dynamical evolution, at the level of the density matrix, can occur in its off-diagonal terms, allowing us to focus the analysis exclusively on quantum coherence.

As pointed out earlier, the pseudomode description of the system offers an insightful point of view on the resonance mechanism, which we can achieve by looking at how the behavior changes with different values of $\lambda$. 
At resonance conditions, the tunneling frequency is such that $J \approx \omega_0$, meaning that the energy levels of the pseudomode oscillators representing the environment(s) are approximately identical to the energy gap between the two single-particle eigenstates $\ket{\Xi^{\pm}}$ of the system Hamiltonian $\hat{H}_S$. Additionally, we are assuming that the full environment starts in the vacuum state $\ket{00_{\text{env}}}$, which is its own ground state. This means that no energy can leak from the environment to the system at the initial time.

\begin{figure}
    \centering
    \includegraphics[width=1\linewidth]{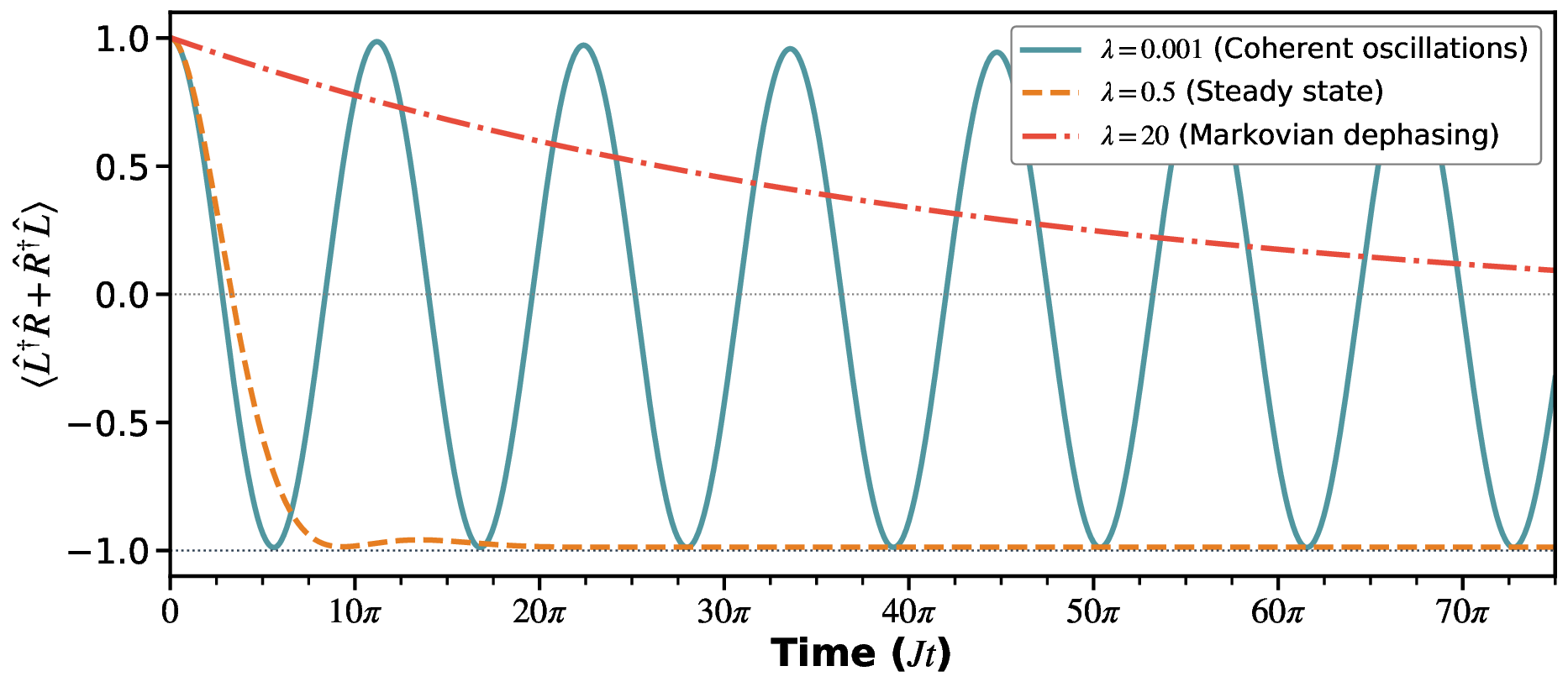}
    \caption{Time evolution of the average value of $\langle \hat{L}^{\dagger}\hat{R} + \hat{R}^{\dagger}\hat{L} \rangle$(t) for the initial state $\ket{\Xi^{\pm}} = \frac{1}{\sqrt{2}}(\ket{L} \pm \ket{R})$, on resonance, for three different values of the coupling distribution width $\lambda$. Blue line: $\lambda = 0.001$. Orange dashed line: $\lambda = 0.5$. Red dashed line: $\lambda = 20$. Other parameters are: $g_0 = 0.1, J= \omega_0 = 2.5, \omega_s = 1, \hbar = 1$. All three curves have been computed through the pseudomode evolution of the reduced system.}
    \label{fig: radiance}
\end{figure}

In the limit of very small dissipation of the pseudomode (i.e. extremely narrow coupling distribution width, $\lambda \rightarrow 0$), the joint system-environment ensemble is free to periodically oscillate between the states $\ket{\Xi^+, 00_{\text{env}}} \leftrightarrow \ket{\Xi^-, 01_\text{env}},\ket{\Xi^-, 10_\text{env}}$, whereas, for strict energy conservation, the initial state $\ket{\Xi^-, 00_\text{env}}$ represents the only ground state, acting like a dark state \cite{scully1997quantum} of the full dynamics. 
As $\lambda$ increases ($\lambda \approx g_0$), the two pseudomodes can significantly lose energy to their dissipative environments, making them incapable of returning the exchanged excitation to the system. Thus, an effective attractor dynamics for the reduced system emerges, where the role of the fixed point is covered by $\ket{\Xi^-}$. Finally, in the limit of very large $\lambda$, the pseudomode is strongly dissipated and the system-environment coupling distribution becomes increasingly flat, thus retrieving the non-Markovian dynamics characterized by pure phase diffusion. The steady state, under such detrimental conditions, becomes an equal mixed combination of $\ket{10}\bra{10}$ and $\ket{01}\bra{01}$, representing complete decoherence. No sharp transitions between the different regimes are observed.
This mechanism is illustrated in Fig.~\ref{fig: radiance}, where the dynamics of the coherence, starting from the higher energy state $\ket{\Xi^+}$, is evaluated in these three regimes. 
As a final remark, we underline that further (non reported) analysis shows that the master equation is able to reproduce both the steady-state attractor dynamics and the Markovian limit, but fails to capture the periodic oscillations. This is expected, as they represent a strong coupling scenario where the approximations under which the master equation has been derived are no longer valid.

\section{Opposite-pseudospin bosons}
\label{sec: opposite}
We consider two identical bosonic qubits initially prepared in opposite pseudospin states and localized in separate spatial regions L and R. The initial pure separable state is
\begin{equation}
    \ket{\Psi(0)}=\ket{L\,\uparrow,R\,\downarrow}.
\end{equation}
We start by outlining the entanglement generation protocol in noiseless conditions. 
The first step is to apply a spatial deformation to the initial state via a unitary transformation, generated by the Hamiltonian $\hat{H}_D$ presented in Eq.~\eqref{eq: deformation_hamiltonian}. The deformation $J t$ is controlled by the tunneling rate $J$ and by the evolution time $t$. When $Jt = n\frac{\pi}{2}$ with $n \in \mathbb{N}$, the spatial wavefunctions become symmetrically distributed among the two spatial regions.
After this process, a post-selective measurement $\hat{\Pi}_{\mathrm{LR}}^{(2)}$ is performed on the final state $\hat{\varrho}_S(t)$ via 
\begin{equation}
    \hat{\Pi}_{\mathrm{LR}}^{(2)} = \sum_{\sigma,\tau=\uparrow,\downarrow} \ket{L\sigma, R\tau}\bra{L\sigma, R\tau}.
\end{equation}
This projector selects only those detection events in which exactly one particle is found in region $L$ and one in region $R$, discarding cases where both particles are detected in the same region. The final normalized state becomes
\begin{equation}
    \hat{\varrho}_{\mathrm{LR}} = \frac{\hat{\Pi}_{\mathrm{LR}}^{(2)} \hat{\varrho}_S(t) \hat{\Pi}_{\mathrm{LR}}^{(2)}}{\mathrm{Tr}(\hat{\Pi}_{\mathrm{LR}}^{(2)} \hat{\varrho}_S(t))}.
\end{equation}
For bosons at $\theta = \pi/2$, the tunneling dynamics produces the following pure state
\begin{equation}
    \begin{split}
    \ket{\Psi(t)}& = \frac{\textbf{i}}{2}(\ket{L\,\uparrow,L\,\downarrow} +\ket{R\,\uparrow,R\,\downarrow})\\
    & + \frac{1}{2}\left(\ket{L{\uparrow}, R{\downarrow}} - \ket{L{\downarrow}, R{\uparrow}}\right).
    \end{split}
\end{equation}
After post-selection, this protocol generates the maximally entangled singlet state
\begin{equation}
    |\Psi_{-}\rangle = \frac{1}{\sqrt{2}}\left(\ket{L{\uparrow}, R{\downarrow}} - \ket{L{\downarrow}, R{\uparrow}}\right),
\end{equation}
with success probability $P_{\mathrm{LR}} = 1/2$. The entanglement is quantified via the concurrence \cite{concurrence} which can be computed as $C(\hat{\varrho}_{\mathrm{LR}}) = \max\left\{0,2\lambda_\mathrm{max} - \text{Tr}(M) \right\},$ where $M = \sqrt{\sqrt{\hat{\varrho}_{\mathrm{LR}}}\Tilde{\varrho}_\mathrm{LR}\sqrt{\hat{\varrho}_{\mathrm{LR}}}}$. The greatest eigenvalue of $M$ is denoted as $\lambda_\mathrm{max}$, while $\Tilde{\varrho}_{\mathrm{LR}} = (\sigma_y \otimes \sigma_y)\hat{\varrho}_{\mathrm{LR}}^{*}(\sigma_y \otimes \sigma_y)$. This entanglement measure reaches unity ($C(\hat{\varrho}_{\mathrm{LR}})=1$) in the ideal case of noiseless spatial deformation \cite{lo2018indistinguishability,nosrati2020robust,nosrati2024indistinguishability}, demonstrating the effectiveness of the protocol. Now, we study the consequence of dephasing noise on the final result of this entanglement generation scheme.
\begin{figure}[t!]
    \centering
    \includegraphics[width=0.99\linewidth]{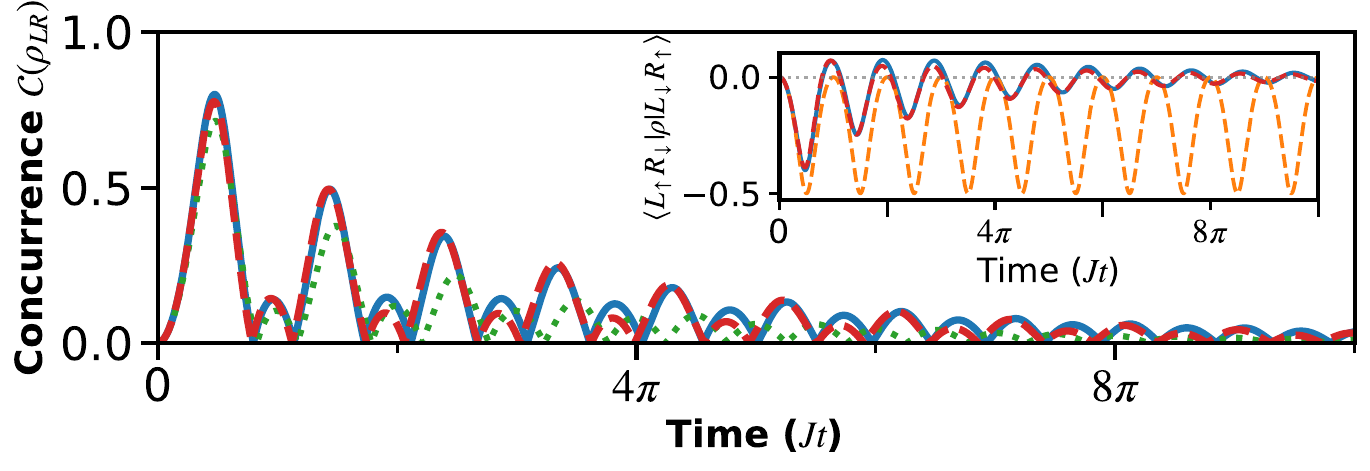}
    \caption{Off-resonance dynamics of entanglement, quantified by concurrence, for opposite-pseudospin bosons from numerical simulations using the full master equation (solid blue line), the phenomenological master equation (dotted green line), and the pseudomode evolution (dashed red line). The inset displays the evolution of the off-diagonal element $\bra{L\,\uparrow,R\,\downarrow}\hat{\varrho}_S(t)\ket{R\,\uparrow,L\,\downarrow}$, where the noiseless case is also shown (dashed orange line). Parameters: $g_0=0.5 $, $\lambda=1.8$, $J =2.5$, $\omega_0 = 0$, $\omega_s=1$, $\hbar = 1$.}
    \label{fig: OFFRES_ent}
\end{figure}
In the off-resonance regime, Fig.~\ref{fig: OFFRES_ent} shows that spin entanglement (quantified by the concurrence) is generated in the vicinity of $Jt = n\pi$ with $n \in \mathbb{N}$, whereas without noise the two particles would remain in a spatially separated, non-entangled state at those times, retrieving the initial state. Examining the off-diagonal elements of the output density matrix, we observe small and asymmetric oscillations that appear before decaying to zero. This indicates that the combined effect of particle hopping and local dephasing gives rise to quantum correlations, even during the expected decay process. However, these correlations are transient and do not persist in the long-time limit, where the system settles into a classical mixture of the states $\ket{L\:\uparrow,L\:\downarrow}, \ket{L\:\uparrow,R\:\downarrow},\ket{L\:\downarrow,R\:\uparrow}, \ket{R\:\uparrow,R\:\downarrow}$, before the action of the post-selective measurement.

\begin{figure}[t!]
    \centering
    \includegraphics[width=0.99\linewidth]{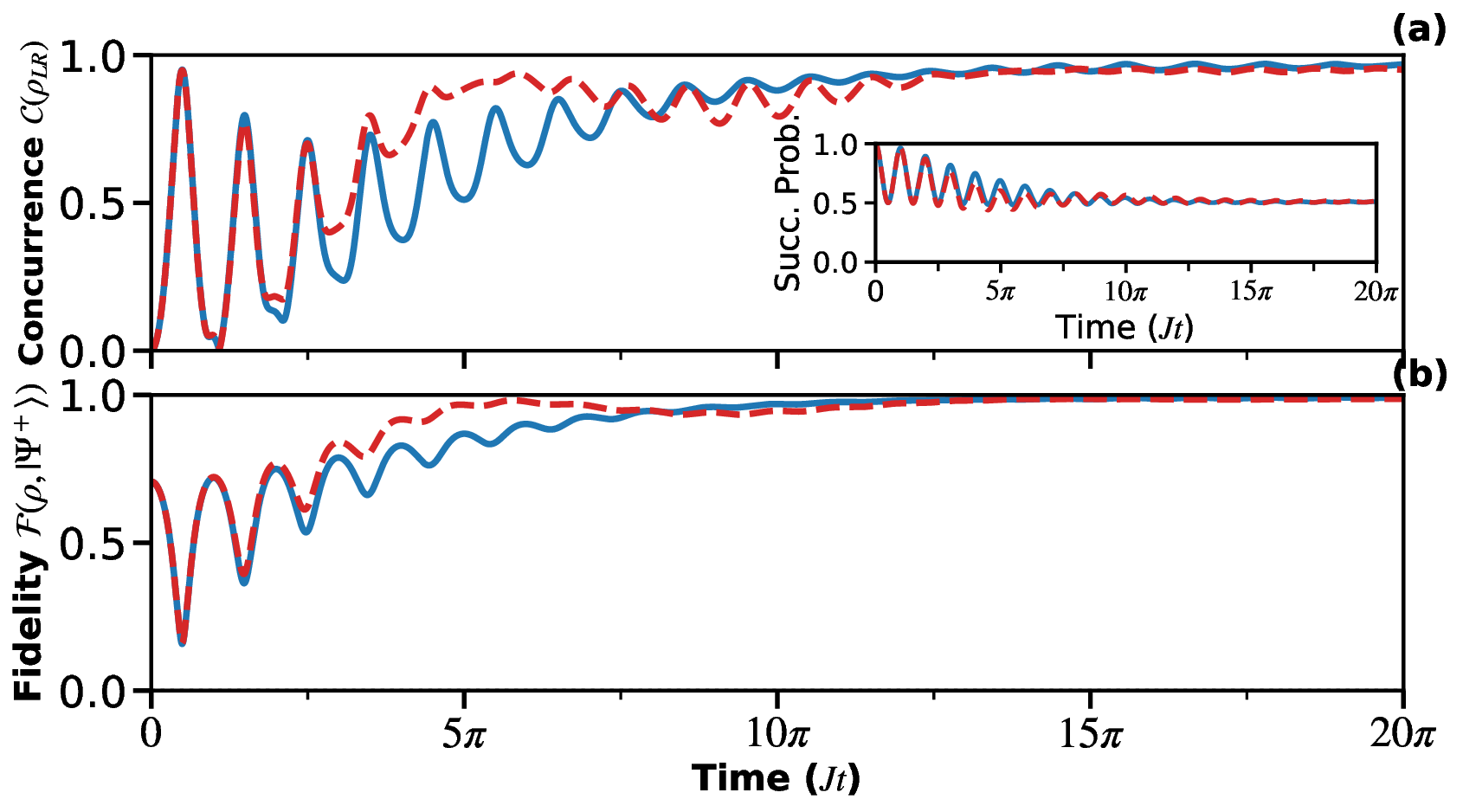}
    \caption{On-resonance dynamics of entanglement, quantified by concurrence, of the distilled state (a) and of fidelity to the Bell state $\ket{\Psi_+}$ (b) for opposite-pseudospin bosons from numerical simulations using the full master equation (solid blue line) and the pseudomode evolution (red dashed line). At the steady state, the trace distance is $\mathcal{T}=0.013790$. Parameters: $g_0 = 0.1, \lambda=0.5, J= \omega_0 = 2.5, \omega_s = 1, \hbar = 1$.}
    \label{fig: OnRES_ent}
\end{figure}

On resonance ($\omega_0 \approx J$), instead, we find that the steady state is entangled, as seen in Fig.~\ref{fig: OnRES_ent}(a). Although in the noiseless case the distillation protocol produces a singlet Bell state $\ket{\Psi_-}$ \cite{nosrati2024indistinguishability}, the steady state emerging here from the joint action of particle tunneling and local dephasing, after post-selection, shows a high fidelity with the Bell triplet state \(
\ket{\Psi_+}=\frac{1}{\sqrt{2}}\big(\ket{L\,\uparrow,R\,\downarrow}+\ket{L\,\downarrow,R\,\uparrow}\big)\), given by \(\mathcal{F}=\bra{\Psi_+}\hat{\varrho}_{\mathrm{LR}}(t)\ket{\Psi_+}\), as displayed in Fig.~\ref{fig: OnRES_ent}(b). This signals how dephasing and coherent tunneling are resonantly generating a different entangled state. 
The reason why the joint system-environment evolution brings about a phase inversion can be understood in light of a similar yet more complicated process as the one described for the single particle case in Sec.~\ref{sec: C}. 
To build the intuition behind this phenomenon, we start by describing some properties of the physical processes under study. 
\begin{itemize}
    \item When we consider the noiseless dynamics ($g_0 = 0$), the singlet and the triplet states evolve in a drastically different way under the bare Hamiltonian $\hat{H}_S$. While the singlet $\ket{\Psi_-}$ is a dark state for the tunneling Hamiltonian (i.e. $\hat{H}_D \ket{\Psi_-} = 0$) and thus it does not change over time, the triplet state has a non-trivial evolution, continuously oscillating between $\ket{\Psi_+}$ and a bunched state characterized by a coherent superposition $\ket{B} \coloneqq \ket{L\,\uparrow,L\,\downarrow} + \ket{R\,\uparrow,R\,\downarrow}$.
    \item Conversely, when we study the sole action of the system-environment interaction with the bath central frequency (which can be interpreted as the $\lambda \rightarrow0$ limit) without tunneling dynamics ($J = 0$), two relevant details emerge. First, the bunched state $\ket{B}$ mentioned above is a dark state for the system-environment interaction, since the net number of spin, measured by $\hat{n}_{\uparrow} - \hat{n}_{\downarrow}$ is identically zero on both spatial regions. Second, under such interaction, the internal coherence of the singlet and the triplet oscillates between its original value and zero. This is because $\hat{H}_{SE}$, as shown in Eq.~\ref{eq: system-bath_hamiltonian}, induces a conditional displacement on the two local environments, displacing their initial vacuum state into opposite coherent states \cite{breuer2002theory} $\ket{\pm \alpha}$, according to the spin of the particle in that specific spatial region. As a result, the internal coherence of the two Bell states is modulated over time via the overlap between the environment states $\langle\alpha|-\alpha\rangle$, which results in a real, positive number. In other words, both Bell states oscillate between a maximally mixed state in the $\{ \ket{L{\uparrow}, R{\downarrow}}, \ket{L{\downarrow}, R{\uparrow}}\}$ basis and a maximally entangled one.
\end{itemize}
Knowing this, the reduced system evolution for the initial state $\ket{\Psi(0)}$, with a finite value of the spectral width $\lambda$ can be described as follows. Initially, the bare Hamiltonian $\hat{H}_S$ tends to evolve the pure, spatially separated state into a superposition of $\ket{\Psi_-}$ and $\ket{B}$. Both of these components fuel the emergence of the triplet state, as a consequence of two distinct but simultaneous processes. On one hand, the singlet state, unacted upon by the system Hamiltonian, loses its internal coherence because of the resonant dephasing, evolving into a mixture of $\ket{\Psi_+}\bra{\Psi_+} + \ket{\Psi_-}\bra{\Psi_-}$. At the same time, the bunched superposition $\ket{B}$, which is not affected by the resonant environments, generates a triplet state $\ket{\Psi_+}$ and cancels out the residual singlet component. Due to the finite size of the coupling distribution, the oscillatory nature of this process is progressively damped until the reduced system reaches a steady state, characterized by a combination of the triplet state $\ket{\Psi_+}$ and a coherent bunched superposition $\ket{B}$. Finally, the action of post-selection activates a specific form of operationally-useful entanglement, extracting the Bell state component with a finite probability of success, as shown in Fig.~\ref{fig: OnRES_ent}.
Due to the non-trivial evolution of the reduced system, we underline that post-selection is not required for the emergence of entanglement in the long-time regime. It is, however, an additional ingredient whose relevance is twofold: on one hand, it directly relates a non-trivial open quantum system dynamics with an experimentally tested framework. On the other hand, it offers extra insights on the structure of the entanglement itself, by projecting the steady state onto the same Hilbert space of the well-known Bell states \cite{piccolini2024asymptotically}, allowing for a direct comparison through the fidelity.
Overall, the newly derived master equation is precise in capturing the long time steady state of the dynamics even when the coupling is relatively strong $\lambda/g_0 \lesssim 0.5$, a regime associated with non-Markovian dynamics \cite{breuer2002theory,bellomo2007PRL,Rivas_2014}. 

\section{Conclusion}
\label{sec: Concl}
We have analytically derived the master equation from the microscopic model of a tunneling system of two bosonic qubits, coupled to local dephasing baths with generic Lorentzian coupling distributions. The analytical results obtained here can be easily extended to a larger $1D$ chain with arbitrary population. Remarkably, we have found that when the main frequencies of the bath are close to the tunneling frequency, the competition between internal coherent dynamics and decoherence gives rise to entangled steady states. As a prospect, our analysis can be extended to bosonic baths with nonzero temperatures. Also, the case of a common dephasing environment together with specific studies of memory effects will be reported elsewhere. 

The tunneling-dephasing interplay analyzed here may be tested in current photonic \cite{bouchard2020two,Wang2020NatPhot} and ultracold atom \cite{cazalilla2011one,Schafer2020NatRev} platforms. In integrated photonic circuits, two-mode bosonic qubits can be encoded in polarization or path degrees of freedom, with local dephasing controllably induced through phase noise or engineered loss channels. Alternatively, ultracold atoms in double-well optical lattices provide a natural realization of the bosonic tunneling Hamiltonian, where local dephasing can be realized via site-dependent light scattering or controlled coupling to tailored reservoirs. In both cases, the resonant regime could be achieved by tuning either the bath spectrum or the tunneling rate. 

We highlight that our model is equivalent to a Bose-Hubbard model with negligible on-site interaction. Thus, further studies on the full Bose-Hubbard model and its fermionic counterpart with specifically engineered dephasing might bring useful insights into the non-trivial interplay between repulsion, tunneling, and noise, specifically in the superfluid phase \cite{kiely2022superfluidity}. 

Our results provide a microscopic foundation for revisiting dephasing models in bosonic tunneling systems and pave the way to general strategies for engineering environment-induced quantum correlations in open many-body platforms.

\begin{acknowledgements}
R.L.F. acknowledges support by MUR (Ministero dell’Università e della Ricerca) through the following projects: PNRR Project ICON-Q -- Partenariato Esteso NQSTI -- PE00000023 -- Spoke 2 -- CUP: J13C22000680006, PNRR Project QUANTIP -- Partenariato Esteso NQSTI -- PE00000023 -- Spoke 9 -- CUP: E63C22002180006, PNRR Project PRISM -- Partenariato Esteso RESTART -- PE00000001 -- Spoke 4 -- CUP: C79J24000190004. A.S. acknowledges support from MUR and Next Generation EU via the PRIN 2022 Project “Quantum Reservoir Computing (QuReCo)” (contract n. 2022FEXLYB) and the NQSTI-Spoke1-BaC project QSynKrono (contract n. PE00000023-QuSynKrono). F.N. acknowledges support by the I+D+i project MADQuantum-CM, financed by the European Union NextGeneration-EU, Madrid Government and by the PRTR.
A.F. would like to thank Matteo Piccolini for a critical reading of the manuscript and Nicola Macrì for useful discussions.
All the codes used for this manuscript are freely available upon reasonable request to the authors.
\end{acknowledgements}
\appendix

\section{Derivation of the master equation}
\label{sec: supp_mat_deriv}
In this section, we show how to derive the master equation for our model.

\subsection{Deriving the exact Interaction Hamiltonian}
We start by defining the Hamiltonian of the system and the environment as 
\begin{equation}
    \hat{H}_{tot} = (\hat{H}_S + \hat{H}_{D}) \otimes \mathbb{I}_{E} + \mathbb{I}_{S} \otimes \hat{H}_{E} +  \hat{H}_{SE}.
\end{equation}
Here, the system Hamiltonian for our bosons is given by two terms, one representing the free evolution and the other introducing a deformation process between the two spatial regions.
\begin{equation}
    \begin{split}
    & \hat{H}_S = \frac{1}{2} \omega_s \sum_{X = \{L, R\}} \sum_{\sigma = \{\uparrow, \downarrow\}} \hat{X}_{\sigma}^{\dagger}\hat{X}_{\sigma}, \\
    & \hat{H}_D = \frac{J}{2} \sum_{\sigma = \{\uparrow, \downarrow\}}\left[ \ld \rsig +  \rd \lsig \right].
    \end{split}
\end{equation}
We stress that $[\hat{H}_S, \hat{H}_D] = 0$. This indicates how the total particle number is conserved by the tunneling process. 
We consider an environment made up by two distinct bosonic baths, which interact with our system through a dephasing process. This can be expressed as
\begin{equation}
    \begin{split}
    & \hat{H}_{E} = \sum_{X = \{L, R \}}\int_0^{\infty} \omega \; \hat{b}_{X}^{\dagger}(\omega) \hat{b}_{X}(\omega) d\omega,\\
    & \hat{H}_{SE} = \sum_{X = \{L, R \}} \left[ \hat{\sigma}_{z, X} \otimes \int_0^{\infty} g_{X} (\omega) [\hat{b}_{X}(\omega) + \hat{b}_{X}^{\dagger}(\omega)] d\omega \right].
    \end{split}
\end{equation}
In this interaction Hamiltonian we have introduced a local dephasing operator $\hat{\sigma}_{z, X} = \hat{n}_{X\uparrow} - \hat{n}_{X\downarrow}$. This operator by itself does not commute with the deformation Hamiltonian $\hat{H}_D$, but it does commute with $\hat{H}_S$ since, as before, this process conserves the total number of particles. 

To derive the master equation, we start by going into interaction picture, thus moving to the frame rotating according to the three contributions $\hat{H}_S + \hat{H}_D + \hat{H}_E$, representing the system evolution and the free environment evolution. To do this, we apply a unitary transformation to the interaction Hamiltonian
\begin{equation}
    \hat{V}(t) = e^{i (\hat{H}_S + \hat{H}_D + \hat{H}_E  )t} \hat{H}_{SE} e^{-i (\hat{H}_S + \hat{H}_D + \hat{H}_E)t}.
\end{equation}
The evolution due to the three different Hamiltonian contributions can be factorized, as each of the three terms commute with the other two. This implies that we can compute the three transformations one by one.
Since $[\hat{H}_S, \hat{\sigma}_{z, X}] = 0$, it follows that $e^{i \hat{H}_{S} t} \hat{\sigma}_{z, X} e^{-i \hat{H}_{S} t} = \hat{\sigma}_{z, X}$.
The environment Hamiltonian brings forth a phase to the creation and annihilation operators of the bath
\begin{equation}
    \begin{split}
          & e^{i \hat{H}_{E} t} H_{SE} e^{-i \hat{H}_{E} t} = \\
          & = \sum_{X = \{L, R \}} \hat{\sigma}_{z, X} \otimes \int g_{X} (\omega) [\hat{b}_{X}(\omega)e^{- i \omega t} + \hat{b}_{X}^{\dagger}(\omega)e^{+ i \omega t}] d\omega \\
          & = \sum_{X = \{L, R \}} \hat{\sigma}_{z, X} \otimes \hat{\mathcal{E}}_X (t),
    \end{split}
\end{equation}
leaving the system part of the operator unaffected. Having this, we can compute the evolution due to the deformation Hamiltonian. To explicitly compute this contribution, we make use of the Baker-Campbell-Hausdorff (BCH) formula \cite{scully1997quantum}, leading us to
\begin{equation}
\begin{split}
    e^{i \hat{H}_{D} t} \hat{H}_{SE} e^{-i \hat{H}_{D} t} & = \hat{H}_{SE} + it [\hat{H}_{D}, \hat{H}_{SE}] \\
    & + \frac{(it)^2}{2!}[\hat{H}_{D}, [\hat{H}_{D}, \hat{H}_{SE}]] + \ldots
     \end{split}
\end{equation}
It is clear that $\el$ and $\er$ will factor out as $\hat{H}_D$ only acts on the system. We compute the commutators one by one
\begin{alignb}
        & [\hat{H}_{D}, \szl] = \frac{J}{2} \left( + \rupc\lupd - \lupc\rupd - \rdownc\ldownd + \ldownc\rdownd \right) := \frac{J}{2} \al,\\
        & [\hat{H}_{D}, [\hat{H}_{D}, \szl ]] = \left(\frac{J}{2}\right)^2 2( \szl -  \szr),\\
        & [\hat{H}_{D}, [\hat{H}_{D}, [\hat{H}_{D}, \szl ]]] = \left(\frac{J}{2}\right)^3 4 \al,\\
        & [\hat{H}_D, [\hat{H}_{D}, [\hat{H}_{D}, [\hat{H}_{D}, \szl ]]]] = \left(\frac{J}{2}\right)^4 8(\szl - \szr).
\end{alignb}
Equivalent results are obtained for $\szr$, by switching indexes $L$ and $R$, noting that $\al = - \ar$. 
We are thus left with the following system-environment Hamiltonian in the interaction picture
\begin{widetext}
\begin{equation}
    \begin{split}
         \hat{V}(t) = & \left[ \frac{(\szl + \szr)}{2}  + \frac{(\szl - \szr)}{2} \cos{(J t)} + \frac{i \al}{2} \sin{(J t)}\right]\otimes\el\\
        &+ \left[ \frac{(\szl + \szr)}{2} + \frac{(\szr - \szl)}{2} \cos{(J t)} + \frac{i \ar}{2} \sin{(J t)}\right]\otimes\er.
    \end{split}
\end{equation}
\end{widetext}
Interestingly, we see that one single local dephasing environment will affect both spatial region considered.
We can cross check this result by applying the BCH formula to each of the operators in the interaction Hamiltonian, since we can rewrite any unitary transformation as
\begin{equation}
        \mathcal{U}(t) \hat{ \mathcal{O}}_1 \hat{ \mathcal{O}}_2 \mathcal{U}^{\dagger}(t) = \mathcal{U}(t) \hat{ \mathcal{O}}_1 \mathcal{U}^{\dagger}(t) \;\cdot\;\mathcal{U}(t)\hat{ \mathcal{O}}_2 \mathcal{U}^{\dagger}(t).
\end{equation}
Specifically, in our case we use the well-known expansion of the creation and destruction operator upon deformation and free evolution
\begin{equation}
    \begin{split}
         \ld (t) =  & e^{i (\hat{H}_S + \hat{H}_D )t} \ld e^{-i (\hat{H}_S + \hat{H}_D )t} =  \\
        & \left[ \ld \cos{\frac{J t}{2}} + i \rd \sin{\frac{J t}{2}} \right]e^{i \omega_s t}.
    \end{split}
\end{equation}
From this, we can plug the pieces together and determine how the number operator for a given spatial region and a given spin transforms in the interaction picture
\begin{equation}
    \begin{split}
            \ld (t) \lsig (t) = & \frac{\ld \lsig + \rd \rsig}{2}+ \frac{\ld \lsig - \rd \rsig}{2} \cos{(J t)}  \\ 
            & + i \frac{\rd \lsig - \ld \rsig}{2} \sin{(J t)}.
    \end{split}    
\end{equation}
This alternative approach, in addition to provide a cross-check, also offers a more intuitive picture of the dynamics and clearly shows how the phase $e^{i \omega_s t}$, induced by the free evolution of our bosons, cancels out in the final expression due to the particles being energetically identical.

\subsection{Derivation of the phenomenological master equation in the short-time regime}
Here we will explicitly show that we can recover the widely adopted Lindblad master equation presented in Eq.~(3) of the main text, in the limit for small $J t$ and for a flat, constant coupling distribution. Under this assumption, the interaction Hamiltonian can be expressed as
\begin{equation}
    \hat{V}(t) \stackrel{J t \; \text{small}}{\approx} \szl \otimes \el + \szr \otimes \er.
    \label{eq: approx_deph}
\end{equation}
In this approximation, the environment operators are uniquely coupled with their respective local dephasing operators. Additionally, it is interesting to note that this exact result would be obtained by neglecting that the deformation Hamiltonian and the local dephasing operator do not commute, i.e. by forcing the condition $[\hat{H_D}, \szl] = [\hat{H_D}, \szr] = 0$ in the derivation of the interaction Hamiltonian. Given this expression for $\hat{V}(t)$, we can derive the time evolution of the density matrix in the interaction picture. First of all, we formally integrate the von Neumann equation, and then we plug the result back into the equation itself. This leads us to
\begin{equation}
    \hat{\varrho}^{I}_{tot}(t) = \hat{\varrho}_{tot}(0) - i \int_0^t [\hat{V}(s), \hat{\varrho}^{I}_{tot}(s)] ds.
\end{equation}
and, subsequently
\begin{equation}
    \frac{d}{dt} \hat{\varrho}^{I}_{tot}(t) = -i [\hat{V}(t), \hat{\varrho}^{I}_{tot}(0)] - \int_0^t[\hat{V}(t),[\hat{V}(s), \hat{\varrho}^I_{tot}(s)]]ds.
\end{equation}
While this is still exact, we should integrate over the whole history of the density matrix to actually solve the integral. To avoid this, we plug the formal result of this equation into the original time evolution one more time and we truncate at second order in the interaction Hamiltonian, thus obtaining
\begin{equation}
    \frac{d}{dt} \hat{\varrho}^{I}_{tot}(t) = -i [\hat{V}(t), \hat{\varrho}^{I}_{tot}(0)] - \int_0^t[\hat{V}(t),[\hat{V}(s), \hat{\varrho}^I_{tot}(t)]]ds.
\end{equation}
From this, we can apply the standard procedure \cite{scully1997quantum, breuer2002theory, manzano2020short} to obtain a Lindblad form. We first apply the Born approximation, under which the system and the environment are weakly coupled, so that the total density matrix can be factorized as $\hat{\varrho}^{I}_{tot}(t) = \hat{\varrho}^{I}_S(t) \otimes \hat{\varrho}_B$. Additionally, we assume that the bath is initially in its vacuum state, i.e. $\hat{\varrho}_B = (\ket{0}\ket{0}...)(...\bra{0}\bra{0})$. This implies that the first term in the time evolution vanishes, since $\langle \el(t) \rangle = \langle \er(t) \rangle = 0$. Being the two bath independent, we can also assume that $\langle \el(t) \er (s) \rangle = 0$ at all times. After we trace out the environment degrees of freedom, we get
\begin{equation}
    \begin{split}
        \frac{d}{dt} \hat{\varrho}^{I}_{S}(t) & = \left[ 2 \szl \; \hat{\varrho}_S^I(t) \; \szl  - \{ \szl, \hat{\varrho}^I_S (t)\} \right]\int_0^{t} \langle \el(t) \el (s) \rangle ds \\
        & + \left[ 2 \szr \;  \hat{\varrho}_S^I(t) \; \szr  - \{ \szr, \hat{\varrho}^I_S (t)\} \right]  \int_0^{t} \langle \er(t) \er (s) \rangle ds,
    \end{split}
\end{equation}
where $\langle \el(t) \el(s) \rangle = \langle \er(t) \er(s) \rangle = g_0 e^{i \omega (t - s)}$.
Now we perform the Markov approximation. This approximation assumes that the bath correlation time is much smaller than the typical time scale of the system, allowing us to replace $s$ with $t-s$, and to extend the upper limit of the integral to infinity. We can then solve the bath-integral correlation integrals as
\begin{equation}
 \begin{split}
    \int_0^{t} \langle \er(t) \er (s) \rangle ds & \stackrel{t \to \infty}{\approx}  \int_0^{\infty} g_0 e^{i \omega s} ds \\
  &  = g_0\left[\delta (\omega) + i\text{P.V.}\left(\frac{1}{\omega}\right)\right].
    \end{split}
\end{equation}
It is relevant to note that here we have taken a constant coupling distribution $g_X (\omega) = g_0, \; X=\{L, R\}$, but in this specific case we would have obtained a similar result with a Lorentzian distribution, leading to a constant coefficient that would depend on the specific features of the distribution, namely its center and its width. Clarified this, we go back into Schr{\"o}dinger picture. In order to do this, we apply once again the approximation for small deformations, thus obtaining the following time evolution for the system density matrix
\begin{equation}
    \Dot{\hat{\varrho}}_S^{S}(t) = -i [\hat{H}, \hat{\varrho}_S^{S}(t)] + g_0\sum_{X = \{L, R\}} D\left[\hat{\sigma}_{z, X} \right]\hat{\varrho}_S^{S}(t).
\end{equation}
Here, we have defined $D[\hat{\mathcal{O}}]\hat{\varrho} = \hat{\mathcal{O}}^{\dagger} \hat{\varrho} \hat{\mathcal{O}} - \frac{1}{2} \{ \hat{\mathcal{O}}^{\dagger} \hat{\mathcal{O}}, \hat{\varrho}\}$. This approximation actually brings us to the Markovian and memoryless master equation shown in the main text in Eq.~(3) of the main text, introducing two local quantum dissipators. However, the phenomena that we are studying, like the HOM effect, do require deformation values for which this approximation might not be reliable ($J t \approx \frac{\pi}{2}$), especially in a strong tunneling regime. Additionally, in a variety of contexts like many-body physics \cite{fazio2024many} long-time dynamics can provide relevant insights over the underlying physics, which might not be caught by a short-time approximation. For these reasons, we derive the full master equation for arbitrary times, which is the subject of the next section.
It is worth mentioning two other fairly recent works on dephasing modeling. \cite{xiong2019boson,kamar2024spin}. In the first one, the dephasing mechanism is described through a cross-Kerr interaction, but no spin is considered. In the second one, the authors consider the same class of system-environment interaction as the present manuscript, but the results are limited to the pure dephasing of a single qubit. Neither of the  two works studies any other dynamical process besides dephasing and free evolution, meaning that there is no spatial dynamics, which is one of the two main physical processes investigated in the present work.

\subsection{Derivation of the master equation for arbitrary times}
In this section we show, step by step, how we derived the full master equation for our system. For simplicity, at this stage we couple only a single spatial region of the system to an ensemble of Bosonic baths. In the next sections, we show how to extend this to both spatial regions.
We start from the previously derived interaction Hamiltonian
\begin{widetext}
\begin{equation}
    \begin{split}
        \hat{V}(t) =  \left[ \frac{(\szl + \szr)}{2}  + \frac{(\szl - \szr)}{2} \cos{(J t)} + i \frac{\al}{2} \sin{(J t)}\right]\otimes\el 
        =  \left[\A + \B \cos{(J t)} + \C \sin{(J t)}\right] \otimes \el.
    \end{split}
\end{equation}
\end{widetext}
Notice that we included $i$ inside the operator $\C$, so that each of the three system operators is Hermitian. Also, $[\A, \B] = [\A, \C] = 0$, while $[\B, \C] = \frac{i}{J} H_D$.  
The environment part of the interaction Hamiltonian is expressed as follows
\begin{equation}
    \el(t)= \int_0^{\infty} g(\omega) [\hat{b}(\omega)e^{- i \omega t} + \hat{b}^{\dagger}(\omega)e^{+ i \omega t}] d\omega.
\end{equation}
We begin by making the two following assumptions:
\begin{equation}
    \begin{split}
        & \hat{\varrho}_B = (\ket{0}\ket{0}...)(...\bra{0}\bra{0}),\\
        & \abs{g(\omega)}^2 = g_0 \frac{1}{\pi} \frac{\lambda}{(\omega - \omega_0)^2 + \lambda^2},
    \end{split}
\end{equation}
that is, the initial state of the bath is the vacuum state, and the distribution of the coupling strengths between the system and the local bath $g(\omega)$ is such that its modulus squared is Lorentzian. This allows to easily compute the bath correlations.   
Following the same procedure as in the previous section, to second order, the evolution of the full density matrix is given by
\begin{equation}
    \frac{d}{dt} \hat{\varrho}^{I}_{tot}(t) = - \int_0^t[\hat{V}(t),[\hat{V}(t), \hat{\varrho}_S(s)\otimes\hat{\varrho}_B]]ds.
\end{equation}
Notice that, once again, we have applied the Born approximation. 
Taking the trace over the environment degrees of freedom and expanding the double commutator, we get
\begin{widetext}
\begin{equation}
    \begin{split}
        \frac{d}{dt}\hat{\varrho}_S^{I}(t) =  - \text{Tr}_{E} \bigg[ \int_0^{t} \hat{V}(t)\hat{V}(s)\hat{\varrho}_S(t)\otimes\hat{\varrho}_B ds 
         - \int_0^{t} \hat{V}(t)\hat{\varrho}_S(t)\otimes\hat{\varrho}_B \hat{V}(s) ds
         - \int_0^{t} \hat{V}(s)\hat{\varrho}_S(t)\otimes\hat{\varrho}_B \hat{V}(t) ds 
         \int_0^{t} \hat{\varrho}_S(t)\otimes\hat{\varrho}_B \hat{V}(s)\hat{V}(t) ds \bigg].
    \end{split}
\end{equation}
\end{widetext}
It is important to point out that here we have applied neither the Markov approximation nor the rotating wave approximation. Such procedure is expected to leave us with a time local, potentially non-Markovian master equation. This equation has sometimes been referred to as the Redfield equation\cite{redfield1957theory, manzano2020short}, although in the literature one can find both Markovian (time independent) and potentially non-Markovian (time dependent) Redfield equations\cite{d2024recovering}. To avoid confusion, we redundantly remark that the present derivation exclusively makes use of a second order perturbative expansion and the Born approximation, keeping a time-local dependence of its correlation coefficients.
Having clarified this, we can split the environment and the system part of the interaction Hamiltonian as $\hat{V}(t)= V_{sys}(t) \otimes \el(t)$, so that, exploiting the cyclicity of the trace, we have
\begin{alignb}
        \frac{d}{dt}\hat{\varrho}_S^{I}(t) = & - \bigg[ \int_0^{t} V_{sys}(t)V_{sys}(s)\hat{\varrho}_S(t) \langle \el(t) \el(s) \rangle ds \\
        & - \int_0^{t} V_{sys}(t)\hat{\varrho}_S(t) V_{sys}(s) \langle \el(s) \el(t) \rangle ds\\
        & - \int_0^{t} V_{sys}(s)\hat{\varrho}_S(t) V_{sys}(t) \langle \el(t) \el(s) \rangle ds \\
        & \int_0^{t} \hat{\varrho}_S(t) V_{sys}(s)V_{sys}(t) \langle \el(s) \el(t) \rangle ds \bigg].
        \label{eq: expansion}
\end{alignb}

To evaluate the bath correlation functions in Eq.~\eqref{eq: expansion}, we exploit the fact that $[ b(\omega_1), b^{\dagger}(\omega_2)] = \delta(\omega_1 - \omega_2)$. Additionally, the environment initial state is the vacuum for each frequency, therefore the only non-zero contribution in the average value is given by the combination term $\langle bb^{\dagger} \rangle = 1$. More specifically, in a vacuum state, we can see that $\langle b^{\dagger} b \rangle = \langle b b \rangle = \langle b^{\dagger} b^{\dagger} \rangle = 0$, so that
\begin{widetext}
\begin{equation}
    \begin{split}
            \langle \el(t) \el(s)\rangle & = 
            \text{Tr}_E \bigg[ \int_0^{\infty} d\omega_1 \int_0^{\infty} d\omega_2 g(\omega_1)g(\omega_2) 
             \times \Big( b(\omega_1) e^{- i\omega_1t}  + b^{\dagger}(\omega_1) e^{ i\omega_1 t}\Big)\Big(b(\omega_2) e^{- i\omega_1 s}  + b^{\dagger}(\omega_2) e^{i\omega_2 s}\Big) \bigg] = \\
            & = \int_0^{\infty} d\omega_1 \int_0^{\infty} d\omega_2 g(\omega_1)g(\omega_2) \langle b(\omega_1) e^{- i\omega_1 t}b^{\dagger}(\omega_2) e^{i\omega_2 s}\rangle \delta(\omega_1 - \omega_2)  \\
            &= \int_0^{\infty} d\omega \abs{g(\omega)}^2 e^{-i \omega (t-s)} = e^{-(i \omega_0 + \lambda) \abs{t - s}}.
    \end{split}
\end{equation}
\end{widetext}
It is worth noting that, with this specific choice of the frequency distributions, the correlation is a complex function of time.

Now, during the derivation of the master equation, we need a few ingredients in the form of various integrals over time of the correlations. We show the building blocks needed for subsequent calculations. To do this, we integrate over the frequencies before we integrate over time:
\begin{equation}
    \begin{split}
           &\int_0^{t} ds e^{-(i\omega_0 + \lambda) \abs{t - s}} =  \alpha(t)e^{-i \omega_0 t - \lambda t}  \frac{e^{i \omega_0 t + \lambda t} - 1}{i \omega_0 + \lambda}, \\
          & \int_0^{t} ds \cos (J s) e^{-(i\omega_0 + \lambda) \abs{t - s}} = \beta(t) = \\
          & =\frac{1}{2} e^{-i \omega_0 t - \lambda t} \left[ \frac{e^{i (\omega_0 + J) t + \lambda t} - 1}{i (\omega_0 + J) + \lambda} + \frac{e^{i (\omega_0 - J) t + \lambda t} - 1}{i (\omega_0 - J) + \lambda} \right],\\ 
           & \int_0^{t} ds \sin (J s) e^{-(i\omega_0 + \lambda) \abs{t - s}} = \kappa(t) = \\
           & = \frac{1}{2i} e^{-i \omega_0 t - \lambda t} \left[ \frac{e^{i (\omega_0 + J) t + \lambda t} - 1}{i (\omega_0 + J) + \lambda} - \frac{e^{i (\omega_0 - J) t + \lambda t} - 1}{i (\omega_0 - J) + \lambda} \right].
    \end{split}
\end{equation}
For the specific case of $\omega_0 = 0$ we get real coefficients
\begin{equation}
    \begin{split}
            & \alpha(t) = \frac{1 - e^{-\lambda t}}{2 \lambda}, \\
            & \beta(t) = \frac{1}{\lambda^2 + J^2}(\lambda \cos{ J t} + J \sin{(J t)} - \lambda e^{- \lambda t}),\\ 
            & \kappa(t) = \frac{1}{\lambda^2 + J^2}(\lambda \sin{ J t} - J \cos{(J t)} + J e^{- \lambda t}).
    \end{split}
\end{equation}

Now we can compute the contributions to the master equation one by one. We have three operators in the interaction Hamiltonian, and when we expand the double commutators in the master equation, we will get terms that are quadratic in the interaction Hamiltonian (namely, $\hat{V}(t)\hat{V}(s)\hat{\varrho}, \hat{\varrho} \hat{V}(s)\hat{V}(t)$, $\hat{V}(t)\hat{\varrho} \hat{V}(s)$ and $\hat{V}(s) \hat{\varrho} \hat{V}(t)$). As a consequence, we will have six possible combinations of the system operators: $\A\A, \B\B, \C\C, \A\B, \A\C, \B\C$, each of them multiplied by a time-dependent coefficient. Since the double commutator brings four different terms, we will have a large number of contributions. 
We can collect all the contributions to the master equation 
\begin{equation}
    \begin{split}
        & \frac{d}{dt} \hat{\varrho}_s^{I}(t) = \\ 
        & g_0\sum_{i, j} \left[G_{i j} \hat{F}_i \hat{\varrho} \hat{F}_j + G_{i j}^{*} \hat{F}_j\hat{\varrho} \hat{F}_i - G_{i j} \hat{F}_j \hat{F}_i \hat{\varrho} - G_{i j}^{*} \hat{\varrho} \hat{F}_i \hat{F}_j\right],
    \end{split}
\end{equation}
with the following complex coefficients matrix:
\begin{equation}
    \mathbb{G} = \begin{pmatrix}
    \alpha(t) & \alpha(t) \cos(J t) & \alpha(t) \sin(J t)\\
    \beta(t)  & \beta(t) \cos(J t) & \beta(t) \sin(J t)\\
    \kappa(t) & \kappa(t) \cos (J t) & \kappa(t) \sin (J t) \\
\end{pmatrix}.
\end{equation}
If we split the real and imaginary parts of the coefficients, we can write it as
\begin{equation}
    \begin{split}
        \frac{d}{dt} \hat{\varrho}_s^{I}(t) = & g_0\sum_{i, j} \Re[G_{i j}] \left[ \hat{F}_i \hat{\varrho} \hat{F}_j + \hat{F}_j\hat{\varrho} \hat{F}_i - \hat{F}_j \hat{F}_i \hat{\varrho} - \hat{\varrho} \hat{F}_i \hat{F}_j\right] \\
        & + i g_0\sum_{i, j}\Im[G_{i j}] \left[ \hat{F}_i \hat{\varrho} \hat{F}_j - \hat{F}_j\hat{\varrho} \hat{F}_i - \hat{F}_j \hat{F}_i \hat{\varrho} + \hat{\varrho} \hat{F}_i \hat{F}_j\right].
    \end{split}
    \label{eq: non_canonical_form}
\end{equation}
This result is general and is valid for any initial state with an arbitrary number of bosons over the two spatial regions, even in superposition states. In the next section, we show how to rewrite the master equation in canonical form, which is useful to analyze the properties of the dynamical map.

\subsection{Canonical form}
Given a generic master equation for the time evolution of the reduced system, it is not always easy to determine whether such evolution describes an actual physical process. To gain some insights on the physicality and have access to widely adopted non-Markovianity measures, it is useful to rewrite the map in canonical form \cite{gorini1976completely, lindblad1976generators, hall2014canonical}, which is given by:
\begin{equation}
    \dot{\hat{\varrho}} = -i [\hat{H}_{\text{can}}(t), \hat{\varrho} ] + \sum_{i} \gamma_{i}(t) D[\hat{\mathcal{O}}_i(t)]\hat{\varrho}.
    \label{eq: canonical_form}
\end{equation}
More specifically, we know that:
\begin{itemize}
    \item If and only if the master equation can be written in canonical form as in Eq.~(\ref{eq: canonical_form}), the resulting dynamical map preserves the trace and the hermiticity of the density matrix. 
    \item If the coefficients  do not depend on time, the map is completely positive (CP) if and only if they are positive; in the case of time-dependent coefficients $\gamma_i (t)$ their positivity is sufficient but no longer necessary for complete positivity.
\end{itemize}
More details can be found in the aforementioned works.

Here we show the mathematical steps to rewrite the master equation into a canonical form and we show that the coefficients can be obtained as the eigenvalues of a specific time dependent, complex coefficients matrix.
We start by rewriting Eq.~(\ref{eq: non_canonical_form}) as:
\begin{equation}
    \begin{split}
        & \frac{d}{dt} \hat{\varrho}_s^{I}(t) = \\ 
        & g_0\sum_{i, j} \left[G_{i j} \hat{F}_i \hat{\varrho} \hat{F}_j + G_{i j}^{*} \hat{F}_j\hat{\varrho} \hat{F}_i - G_{i j} \hat{F}_j \hat{F}_i \hat{\varrho} - G_{i j}^{*} \hat{\varrho} \hat{F}_i \hat{F}_j\right].
    \end{split}
\end{equation}
First, we focus on terms with the same index $i=j$.
\begin{equation}
    \begin{split}
        & \sim g_0 \sum_{i} \left[ G_{ii} \hat{F}_i \hat{\varrho} \hat{F}_i + G_{ii}^{*} \hat{F}_i \hat{\varrho} \hat{F}_i - G_{ii} \hat{F}_i \hat{F}_i \hat{\varrho} - G_{ii}^{*} \hat{\varrho} \hat{F}_i \hat{F}_i\right] \\
        & =  g_0 \sum_{i} \left[2 \Re{(G_{ii})} \hat{F}_i \hat{\varrho} \hat{F}_i - \Re{(G_{ii})} \hat{\varrho} \hat{F}_i \hat{F}_i - \Re{(G_{ii})} \hat{F}_i \hat{F}_i \hat{\varrho} \right]\\
        & - i \left[ \Im{(G_{ii})} \hat{F}_i \hat{F}_i \hat{\varrho} - \Im{(G_{ii})} \hat{\varrho} \hat{F}_i \hat{F}_i + \right] \\
        & = \sum_{i} \left[ 2 \Gamma_{ii} D[\hat{F}_i]\hat{\varrho} - i g_0 \Im{(G_{ii})} [\hat{F}_i^2, \hat{\varrho}] \right].
    \end{split}
\end{equation} 
Here we have used the definition of the dissipator $D[\hat{F}_i]\hat{\varrho}$ and we have defined $\Gamma_{ii} = 2 g_0 \Re{(G_{ii})}$. This shows that the diagonal terms in the coefficients matrix $\mathbb{G}$ can be easily rearranged into a term proportional to $D[\hat{F}_i]$ and a hermitian Hamiltonian part $\sim g_0 G_{ii} \hat{F}_i^2$.
Now we focus on the off-diagonal terms, i.e., $i \neq j$. Specifically, we first group the terms with $\sim \hat{F}_i \hat{\varrho} \hat{F}_j$, namely
\begin{equation}
    \begin{split}
        & \sim g_0 \sum_{i \neq j} \left[ G_{ij} \hat{F}_i \hat{\varrho} \hat{F}_j + G_{ij}^{*} \hat{F}_j \hat{\varrho} \hat{F}_i\right] = \\
        & = g_0 \sum_{i < j} \left[ (\Re{(G_{ij})} + \Re{(G_{ji})}) (\hat{F}_i \hat{\varrho} \hat{F}_j + \hat{F}_j \hat{\varrho} \hat{F}_i) \right]\\
        & + i g_0 \sum_{i < j} \left[ (\Im{(G_{ij})} - \Im{(G_{ji})}) \hat{F}_i \hat{\varrho} \hat{F}_j \right] \\
        & - i g_0 \sum_{i < j} \left[ (\Im{(G_{ij})} - \Im{(G_{ji})}) \hat{F}_j \hat{\varrho} \hat{F}_i \right] \\
        & = \sum_{i, j} \Gamma_{ij} \hat{F}_i \hat{\varrho} \hat{F}_j,
        \label{eq: first_part_non_diag}
    \end{split}
\end{equation}
where we have defined $\Gamma_{ij} = g_0 (\Re{(G_{ij})} + \Re{(G_{ji})}) + i g_0 (\Im{(G_{ij})} - \Im{(G_{ji})})$. Note that for $i=j$ this definition is consistent with the one we gave before.
Now we focus on the remaining terms, which are of the form $\sim \hat{F}_j \hat{F}_i \hat{\varrho}$ and $\sim \hat{\varrho} \hat{F}_i \hat{F}_j$, that is
\begin{equation}
    \begin{split}
        & \sim g_0\sum_{i \neq j} \left[- G_{ij} \hat{F}_j \hat{F}_i \hat{\varrho} - G_{ij}^{*} \hat{\varrho} \hat{F}_i \hat{F}_j\right] \\
        & = - g_0\sum_{i \neq j} \left[\frac{G_{ij}}{2} \hat{F}_j \hat{F}_i \hat{\varrho} + \frac{G_{ij}}{2} \hat{F}_j \hat{F}_i \hat{\varrho}  \right] \\
        & - g_0\sum_{i \neq j} \left[ + \frac{G_{ij}^{*}}{2} \hat{\varrho} \hat{F}_i \hat{F}_j + \frac{G_{ij}^{*}}{2} \hat{\varrho} \hat{F}_i \hat{F}_j \right] \\
        & = - g_0\sum_{i \neq j} \left[\frac{G_{ij}}{2} \hat{F}_j \hat{F}_i \hat{\varrho} + \frac{G_{ij}}{2} \hat{F}_i \hat{F}_j \hat{\varrho} - \frac{G_{ij}}{2}[\hat{F}_i, \hat{F}_j]\hat{\varrho} \right] \\
        & - g_0\sum_{i \neq j} \left[ + \frac{G_{ij}^{*}}{2} \hat{\varrho} \hat{F}_i \hat{F}_j + \frac{G_{ij}^{*}}{2} \hat{\varrho} \hat{F}_j \hat{F}_i + \frac{G_{ij}^{*}}{2} \hat{\varrho} [\hat{F}_i, \hat{F}_j] \right] \\
        & = + g_0 \sum_{i,j} \Re{(G_{ij})} \left[- \frac{1}{2} \{\hat{F}_j \hat{F}_i, \hat{\varrho}\} - \frac{1}{2} \{\hat{F}_i \hat{F}_j, \hat{\varrho}\} \right] \\
        & + i g_0\sum_{i,j} \Im{(G_{ij})} \left[\frac{1}{2} \{\hat{F}_i \hat{F}_j, \hat{\varrho}\} - \frac{1}{2} \{\hat{F}_j \hat{F}_i, \hat{\varrho}\} + \right] \\
        & - g_0 \sum_{i,j} \frac{\Re{(G_{ij})}}{2} [[\hat{F}_i, \hat{F}_j], \hat{\varrho}] \\
        & - i g_0\sum_{i,j}\frac{\Im{(G_{ij})}}{2} [\{\hat{F}_i, \hat{F}_j\}, \hat{\varrho}] .
    \end{split}
\end{equation}
We can use the previous definition of $\Gamma_{ij}$ and follow the same steps as in Eq.~\eqref{eq: first_part_non_diag} to rewrite the first two lines as $ - \sum_{i,j} \Gamma_{ij} \frac{1}{2} \{\hat{F}_j \hat{F}_i, \hat{\varrho}\}$. The last two lines, instead, can be collected into a Hamiltonian part, since they are of the form $-i [H', \hat{\varrho}]$. 
By rearranging all terms, we get the following non-diagonal canonical form for the master equation:
\begin{equation}
    \dot{\hat{\varrho}} = -i [\hat{H}_{\text{can}}(t), \hat{\varrho} ] + \sum_{i, j} \Gamma_{ij} \left[ \hat{F}_i \hat{\varrho} \hat{F}_j - \frac{1}{2} \{ \hat{\varrho}, \hat{F}_j \hat{F}_i\}\right],
    \label{eq: final_ME}
\end{equation}
where we have defined
\begin{equation}
    \begin{split}
        &  \Gamma_{ij}(t) = g_0 (G_{ij}(t) + G_{ji}(t)^{*}), \\
        & \hat{H}_{\text{can}}(t) = \sum_{i,j} \frac{g_0}{2i}  (G_{ij}(t) - G_{ji}^{*}(t)) \hat{F}_j \hat{F}_i.
    \end{split}
\end{equation}

\subsection{Full master equation for two local identical dephasing baths}
Now that we have solved the case for dephasing noise in a single spatial region, we move on to extend this to two spatial regions. By looking at the full interaction Hamiltonian,
\begin{widetext}
\begin{equation}
    \begin{split}
         \hat{V}(t) = & \left[ \frac{(\szl + \szr)}{2}  + \frac{(\szl - \szr)}{2} \cos{(J t)} + i \frac{\al}{2} \sin{(J t)}\right]\otimes\el +\\
        & \left[ \frac{(\szl + \szr)}{2} + \frac{(\szr - \szl)}{2} \cos{(J t)} + i \frac{\ar}{2} \sin{(J t)}\right]\otimes\er = \\ 
        & \left[ \A + \B \cos{(J t) + \C \sin{(J t)}} \right] \otimes \el         
        + \left[ \A - \B \cos{(J t) - \C \sin{(J t)}} \right] \otimes \er,
    \end{split}
\end{equation}
\end{widetext}
we see that we have some interesting symmetries. Specifically, each of the three system operators is coupled to both environments, but the operators coupled to the second environment have a minus sign in front of $\B$ and $\C$.
As an additional assumption, we consider that the two distinct baths are uncorrelated at the initial time, and that are left uncorrelated by their free dynamics.. This translates into $\langle \el(t) \er(s)\rangle=0, \; \forall s,t$, and is in line with the setting of two local independent baths. Therefore, when expanding the double commutator, the only surviving terms are those connecting the same local environment. 
If we assume that the two distinct environments have the same coupling distribution and they are both initially in a vacuum state, we can see that local dephasing in either spatial region brings to the similar density matrix evolution as in Eq.~\ref{eq: final_ME}. Mathematically, this means that we can extend  Eq.~\ref{eq: final_ME} simply by multiplying both $\hat{H}_{can}(t)$ and each element $\Gamma_{ij}$ by two, and cancel out the minus signs in front of $\B$ and $\C$. Since the operator $\A$ has the same sign in front of both environments, the cross terms $\sim \A\;\B$ and $\sim \A\;\C$ are canceled. 
Therefore, we reach the final master equation for two local identical dephasing baths that we used for the results in the main text, given by
\begin{equation}
    \dot{\hat{\varrho}} = -i [\hat{H}_{\text{can}}(t), \hat{\varrho} ] + \sum_{i, j} \Gamma_{ij} \left[ \hat{F}_i \hat{\varrho} \hat{F}_j - \frac{1}{2} \{ \hat{\varrho}, \hat{F}_j \hat{F}_i\}\right],
    \label{eq: final_ME}
\end{equation}
with
\begin{equation}
    \begin{split}
        &  \Gamma_{ij} = 2g_0 (G_{ij} + G_{ji}^{*}), \\
        & \hat{H}_{\text{can}}(t) = 2 \sum_{i,j} \frac{g_0}{2i}  (G_{ij} - G_{ji}^{*}) \hat{F}_j \hat{F}_i, \\
        & \mathbb{G} = \begin{pmatrix}
    \alpha(t) & 0 & 0\\
    0  & \beta(t) \cos(J t) & \beta(t) \sin(J t)\\
    0 & \kappa(t) \cos (J t) & \kappa(t) \sin (J t) \\
\end{pmatrix}.
    \end{split}
\end{equation}
In principle, we can further simplify the master equation by diagonalizing the $\Gamma$ matrix, so that we can rewrite the master equation in terms of three new jump operators, which are linear combinations of the $\hat{F}_i$. More specifically, we can find a unitary matrix $U$ such that $\Gamma = U \Lambda U^{\dagger}$, where $\Lambda$ is a diagonal matrix containing the eigenvalues of $\Gamma$. Then, by defining the new jump operators as $\hat{\mathcal{O}}_i = \sum_j U_{j}^i \hat{F}_j$, we can rewrite the master equation in a fully diagonal canonical form. The equation for the eigenvalues can be formally expressed as a third order complex coefficients polynomial
\begin{equation}
    \begin{split}
        & \det(\Gamma - \gamma I) = 0 \Rightarrow \\
        & +\gamma^3 - \gamma^2 \text{Tr}(\Gamma) + \frac{\gamma}{2} [\text{Tr}(\Gamma^2) - \text{Tr}(\Gamma)^2] - \det(\Gamma) = 0 \\
        & \Rightarrow \gamma^3  - \gamma^2 (\alpha(t) + \beta(t)\cos(J t) + \kappa(t) \sin(J t)) \\
        & + \gamma \bigg[\beta(t) \kappa(t) \cos(J t) \sin(J t) - \alpha(t) \beta(t) \cos(J t) - \alpha(t) \kappa(t) \sin(J t) \\
        & + \abs{\beta(t) \sin(J t) + \kappa(t) \cos(J t)}^2\bigg] \\
        & - \alpha(t) \abs{\beta(t) \sin(J t)+ \kappa(t) \cos(J t)}^2  \\
        & - \alpha(t)\beta(t) \kappa(t) \cos(J t) \sin(J t) = 0.
    \end{split}
\end{equation}
Finding an explicit analytical solution is not a trivial task since the expressions could be quite cumbersome.
Regardless, we are interested in the properties of the $\Gamma$ matrix, which is Hermitian by construction. Therefore, we can numerically compute its eigenvalues at all times and check whether they are positive or not. We show the results in the following sections.

\section{Physicality and non-Markovianity of the master equation}
\label{sec: non-mark}
In the previous sections, we have seen how we can rewrite the full master equation in a non diagonal form, with
\begin{equation}
    \dot{\hat{\varrho}} = -i [\hat{H}_{\text{can}}, \hat{\varrho} ] + \sum_{i, j} \Gamma_{ij} \left[ \hat{F}_i \hat{\varrho} \hat{F}_j - \frac{1}{2} \{\hat{F}_j \hat{F}_i, \hat{\varrho}\}\right]\text{.}
    \label{eq: non_diag_canonical_form}
\end{equation}
Here, the matrix $\Gamma$ is known as the Kossakowski matrix or diffusion matrix \cite{gorini1976completely}.
It is known that a map written in this form, with a Hermitian $\Gamma$, is trace-preserving and Hermiticity-preserving. This can be easily confirmed by (i) computing the time derivative of the trace of the density matrix which results equal to zero and (ii) computing the Hermitian conjugate of the whole time derivative.
Following \cite{benatti2005open}, we know that, if the Kossakowski matrix is positive definite, the map defined by Eq.~\ref{eq: canonical_form} is Completely Positive and Trace Preserving ($CPTP$). However, in the case of time dependant coefficients, this is no longer a necessary condition. Instead, negative eigenvalues of $\Gamma$ can be a signature of non-Markovianity, according to commonly adopted definitions \cite{rivas2010entanglement, hall2014canonical}. Additionally, in specific cases in which the eigenvalues of $\Gamma$ are time-dependent and one of them is always negative, this can be interpreted as a trademark of a particular phenomenon known as eternal non-Markovianity \cite{megier2017eternal}.

\begin{figure}
    \centering
    \includegraphics[width=0.99\linewidth]{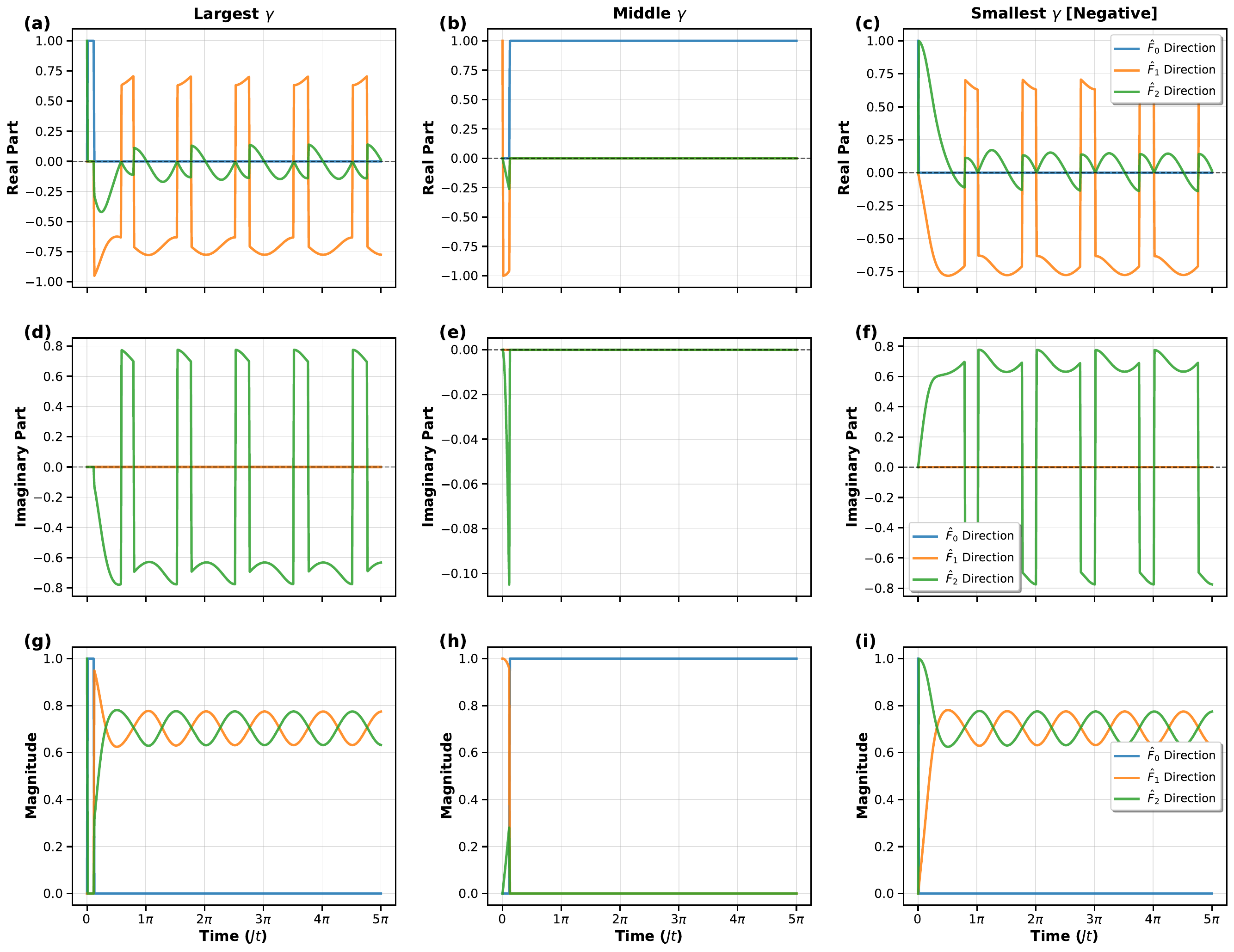}
    \caption{Time evolution of the direction of each eigenvector of the Kossakowski matrix $\Gamma$ in the 3D space spanned by the operators $\A, \B, \C$. The parameters are the same as in Fig.~\ref{fig: eigenvalues}, corresponding to the on-resonance case with $\omega_0 = J$. }
    \label{fig: eigenvectors}
\end{figure}

In Fig.~\ref{fig: eigenvalues}(a), we show the time evolution of the three eigenvalues of $\Gamma$ for a specific set of parameters, corresponding to the on-resonance scenario. We can see that one of them is always negative after the initial time. In Fig.~\ref{fig: eigenvalues}(b), instead, we show the minimum eigenvalue of $\Gamma$ over time for varying values of the average value of the Lorentzian coupling distribution $\omega_0$. We can see that, regardless of the value of $\omega_0$, one of the eigenvalues is always negative (or, at most, equal to zero). This suggests that the map is always non-Markovian over the evolution times considered, which could be associated to emergence of steady state entanglement \cite{huelga2012non}. We remark that a similar behavior is observed for varying values of the other parameters, i.e. $g_0, \lambda, J$. Qualitatively, this phenomenon is stronger for low values of $\lambda$ and high values of $g_0$, i.e. when the environment has a long memory time and the system-environment coupling is strong. This is in line with the common intuition that non-Markovian effects are more relevant in these regimes. To gain more insight on the non-Markovian nature of the map, we can rewrite the master equation in diagonal form
\begin{equation}
    \dot{\hat{\varrho}} = -i [\hat{H}_{\text{can}}, \hat{\varrho} ] + \sum_{i \in \{0, -, +\}} \gamma_i \left[ \hat{\mathcal{O}}_i \hat{\varrho} \hat{\mathcal{O}}^{\dagger}_i - \frac{1}{2} \{\hat{\mathcal{O}}^{\dagger}_i \hat{\mathcal{O}}_i, \hat{\varrho}\}\right]\text{.}
    \label{eq: diag_canonical_form}
\end{equation}
As mentioned before, the new jump operators $\hat{\mathcal{O}}_i$ are linear combinations of the original ones $\hat{\mathcal{O}}_i = \sum_k U_{k}^i \hat{F}_k$, and the new coefficients $\gamma_i$ are the eigenvalues of the Kossakowski matrix $\Gamma$. $U$ is the unitary matrix which diagonalizes $\Gamma$. Also, we remark that such form is unique \cite{hall2014canonical}. By explicitly computing the eigenvectors which correspond to each eigenvalue (i.e. the columns of the matrix $U$, denoted as $\Vec{U}^i$) one can infer more details about the physical meaning of each jump operator. In figure~\ref{fig: eigenvectors}, we show the time evolution of the direction of each eigenvector in the 3D space spanned by the operators $\A, \B, \C$.

Interestingly, we find that one of the eigenvalues has a unit eigenvector which is entirely aligned on the direction of $\A$, except for an initial short time window. We will denote this as $\gamma_0$. Additionally, the other two eigenvectors do not have any relevant component on the aforementioned direction. This shows how, as a consequence, the operator $\A$ is effectively decoupled from the other two: the result is a global dephasing effect. This is an interesting detail, as it shows clearly how the identical, local dephasing we are considering, can be decomposed in an effective global dephasing effect, plus two other contributions which we are now going to consider. 
Specifically, as commented before and illustrated in Fig.~\ref{fig: eigenvalues}, the Kossakowski matrix has two extra non-zero eigenvalues.  One of them, denoted as $\gamma_{+}$, is always positive, while the other one is always negative ($\gamma_{-}$).
Due to the direction of the eigenvectors, we can write the jump operators of the diagonal master equation as
\begin{equation}
    \begin{split}
        & \hat{\mathcal{O}}_0 (t)= \A, \\
        & \hat{\mathcal{O}}_+ (t)= U^+_1 (t) \B + U^+_2 (t) \C, \\
        & \hat{\mathcal{O}}_- (t)= U^-_1 (t) \B + U^-_2 (t) \C,
    \end{split}
\end{equation}
where $U^{\pm}_1(t), U^{\pm}_2(t)$ are time-dependent coefficients determined by the corresponding columns of the unitary matrix $U$ (specifically, of its $2$x$2$ block corresponding to $\B$ and $\C$). Plugging the expressions for the operators $\hat{F}_k$ we get a clearer picture of the physical processes described by these new operators $\hat{\mathcal{O}}_i$
\begin{equation}
    \begin{split}
        & \hat{\mathcal{O}}_0 (t)\approx \sum_{\sigma} \hat{L}_{\sigma}^{\dagger} \hat{L}_{\sigma} + \hat{R}_{\sigma}^{\dagger} \hat{R}_{\sigma}, \\
        & \hat{\mathcal{O}}_+ (t)= \sum_{\sigma} U^+_1 (t) \; [\hat{L}_{\sigma}^{\dagger} \hat{L}_{\sigma} - \hat{R}_{\sigma}^{\dagger} \hat{R}_{\sigma}] + U^+_2 \; (t) i [\hat{R}_{\sigma}^{\dagger} \hat{L}_{\sigma} - \hat{L}_{\sigma}^{\dagger} \hat{R}_{\sigma}],\\
        & \hat{\mathcal{O}}_- (t)= \sum_{\sigma} U^-_1 (t) \; [\hat{L}_{\sigma}^{\dagger} \hat{L}_{\sigma} - \hat{R}_{\sigma}^{\dagger} \hat{R}_{\sigma}] + U^-_2  \; (t) i [\hat{R}_{\sigma}^{\dagger} \hat{L}_{\sigma} - \hat{L}_{\sigma}^{\dagger} \hat{R}].
    \end{split}
    \label{eq: jump_operators_decomposed}
\end{equation}
Here, we can see that the first operator $\hat{\mathcal{O}}_0$ measures the total number of particles in both spatial regions, while the other two operators are linear combinations of (i) the difference in the number of particles between the two regions and (ii) the particle tunneling between the two spatial regions. Therefore, we can interpret the jump operators associated to $\gamma_{+}(t)$ and $\gamma_{-}(t)$ as a combination of spatial dephasing noise and noise affecting the particle hopping between the two regions. This interpretation holds true also in the case of opposite-pseudospin bosons, where each term is summed over both spin components. Note that both $\hat{\mathcal{O}}_- (t)$ and $\hat{\mathcal{O}}_+ (t)$ have the same operatorial structure as the system Hamiltonian $\hat{H}_S + \hat{H}_D$, with a $\frac{\pi}{2}$ phase shift between the two spatial regions, and with oscillating coefficients. Additionally, the time dependence of such coefficients $U^{\pm}_1(t), U^{\pm}_2(t)$, that we computed numerically, suggests how the relative weight of each contribution oscillates in opposite phases, thus bringing to an alternation between Markovian and non-Markovian dynamics. The antiphase oscillations are particularly evident in the magnitude of the coefficients, as suggested by the last row of Fig.~\ref{fig: eigenvectors}.
As an additional signature of information backflow, we can investigate violation of P-divisibility of our dynamical map. Having obtained a diagonal form of the master equation, we can adopt known results to check the P-divisibility \cite{chruscinski2022dynamical, benatti2024quantum}.
Specifically, a master equation in diagonal form is P-divisible if and only if, given the matrix
\begin{equation}\label{WJmatrix}
    W_J^{\ket{\phi}}(t) = \sum_{i \in \{0, -, +\}} \gamma_i (t) \big[ \hat{\mathcal{O}}_i (t) - l_{i, \ket{\phi}} \big]\ket{\phi} \bra{\phi}\big[ \hat{\mathcal{O}}_i (t) - l_{i, \ket{\phi}} \big]^{\dagger},
\end{equation}
its eigenvalues are positive for all time $t$. Here, $l_{i, \ket{\phi}} = \bra{\phi}\hat{\mathcal{O}}_i \ket{\phi}$. Since this is a necessary and sufficient condition, if we find a single state $\ket{\phi}$ for which $W_J^{\ket{\phi}}(t)$ exhibits a negative eigenvalue, we can pinpoint the time interval in which the dynamical map is not P-divisible.
In order to exemplify this, we pick three specific states 
\begin{equation}
    \begin{split}
        & \ket{\Psi_\pm} = \frac{1}{\sqrt{2}} (\ket{L\uparrow,R\downarrow} \pm \ket{L\downarrow,R\uparrow}),\\
        & \ket{\Phi} = \frac{1}{2}(\ket{0}_L - \ket{1}_L)\otimes(\ket{0}_R + \ket{1}_R),
    \end{split}
\end{equation}
and we numerically compute the eigenvalues of $W_J^{\ket{\phi}}(t)$ at all times. Here, we define as $\ket{n}_X$ the state vector representing $n$ equal-spin particles in the $X$ spatial region. The first two states are maximally entangled states over the two spatial regions, while the second one is a product state of local superposition states. This choice allows us to investigate the influence of different kinds of quantum correlations on the P-divisibility analysis. 

\begin{figure}
    \centering
    \includegraphics[width=0.99\linewidth]{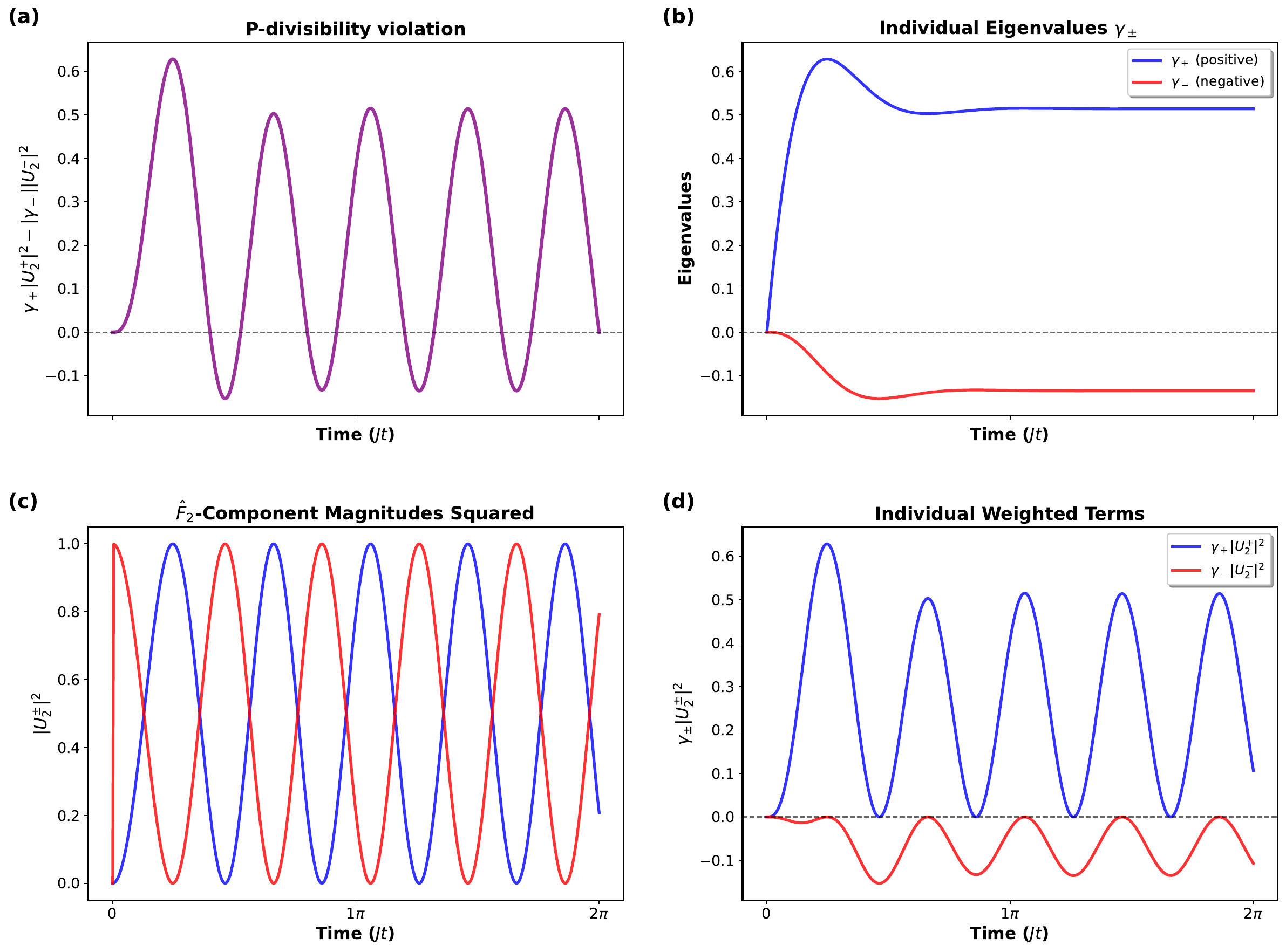}
    \caption{Time evolution of the eigenvalue of $W_J^{\ket{\Psi_+}}$ for the specific choice of the entangled Bell states $\ket{\Psi_+}$, computed from the noise operators in the diagonal form of the master equation shown in Eq.~(\ref{eq: jump_operators_decomposed}), in the case of opposite-pseudospin bosons. The parameters are the same as in Fig.~(2) of the main text, corresponding to the off-resonance case with $\omega_0 = 0$. This shows how p-divisibility breaks can emerge even out of resonance.}
    \label{fig: p_divisibility}
\end{figure}

\begin{figure}
    \centering
    \includegraphics[width=0.99\linewidth]{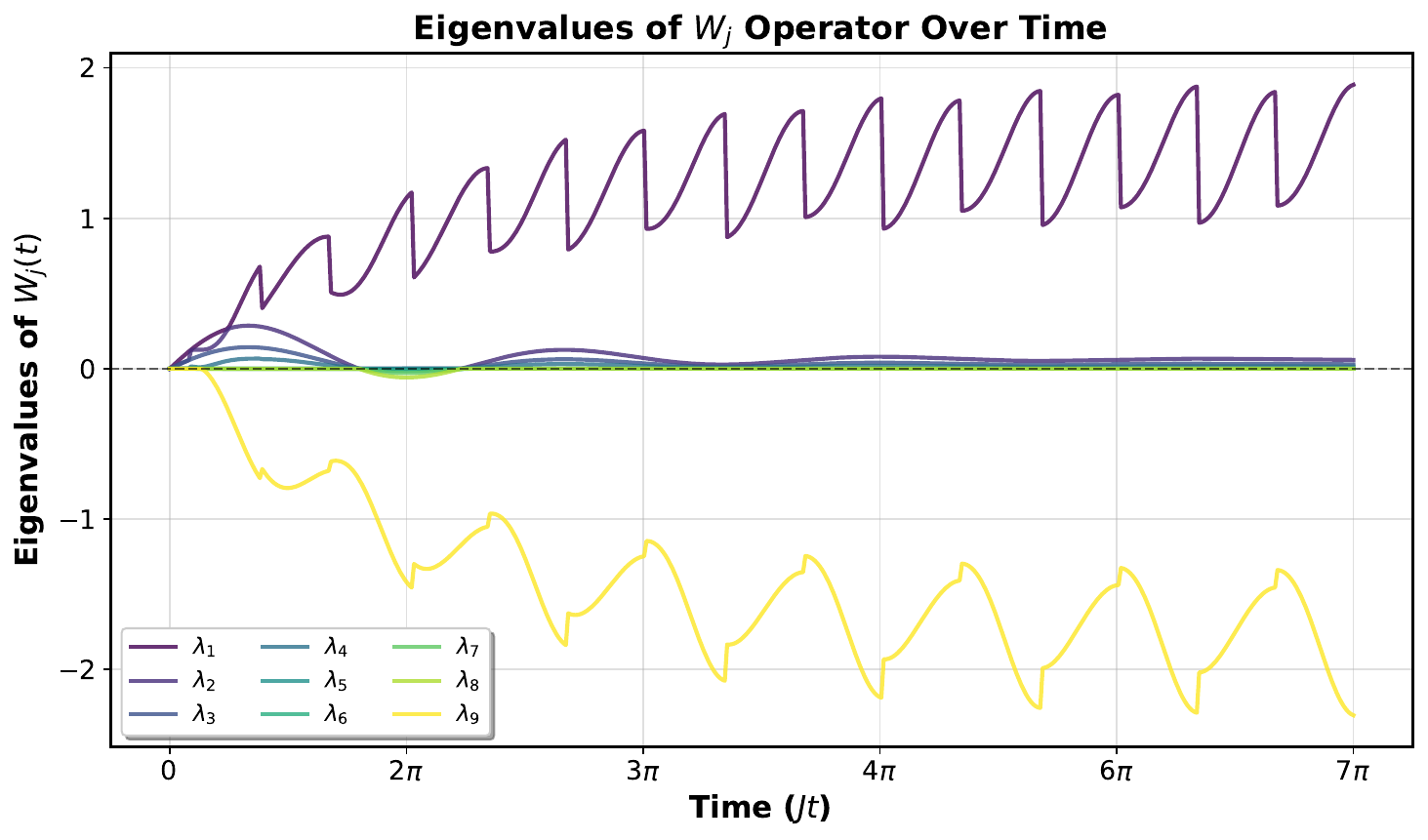}
    \caption{Time evolution of the eigenvalue of $W_J^{\ket{\Phi}}$ for the specific choice of orthogonal states $\ket{\psi} = \frac{1}{2}(\ket{0}_L - \ket{1}_L)\otimes(\ket{0}_R + \ket{1}_R)$, computed from the noise operators in the diagonal form of the master equation shown in Eq.~\ref{eq: jump_operators_decomposed}, in the case of opposite-pseudospin bosons. The parameters are the same as in Fig.~(5) of the main text, corresponding to the on-resonance case with $\omega_0 = J$.}
    \label{fig: p_div_new}
\end{figure}

We start with the maximally entangled states $\ket{\Psi_\pm}$, because in this case the matrix $W_J^{\ket{\phi}}(t)$ can be analytically computed. The action of the new noise channels on the state itself is given by
\begin{equation}
    \begin{split}
        & \hat{\mathcal{O}}_0 \ket{\Psi_\pm} = \ket{\Psi_\pm}, \\
        & \hat{\mathcal{O}}_{\pm} \ket{\Psi_+} = \frac{1}{\sqrt{2}} i U^{\pm}_2(t) (\ket{R\uparrow,R\downarrow} - \ket{L\uparrow,L\downarrow}), \\
        & \hat{\mathcal{O}}_{\pm} \ket{\Psi_-} = 0.
    \end{split}
\end{equation}
From this expression, we can see that the contribution of the overall dephasing channel $\hat{\mathcal{O}}_0$ on both regions vanishes, while the contributions of the other two channels depend only on the coefficients $U^{\pm}_2(t)$, which weight the tunneling operator in each jump operator. This suggests that for this specific set of states the violation of P-divisibility of the map is mostly linked to noise affecting the particle hopping between the two spatial regions (since the terms proportional to $U^{\pm}_1(t)$ vanish). Additionally, here, the relative phase in the noise operator is crucial to have a non-zero contribution for the triplet state, since it switches the relative phase in the entangled state. This, by itself, is a result of the non-trivial time evolution of the system-environment interaction operators in the interaction picture. 
The resulting matrix can be explicitly computed, leading us to
\begin{equation}
    \begin{split}
        & W_J^{\ket{\Psi_+}}(t) = \left(\abs{U_{2}^+(t)}^2 \gamma_+ (t) - \abs{U^-_2 (t)}^2 \abs{\gamma_-(t)}\right)   \begin{pmatrix}
1 & -1 \\
-1 & 1
\end{pmatrix},
    \end{split}
\end{equation}
where the matrix is written in the restricted space of $\ket{L\uparrow,L\downarrow}$ and $\ket{R\uparrow,R\downarrow}$. 
There is only one nonzero real eigenvalue $\lambda$, which is proportional to the expression written above
\begin{equation}
    \lambda = 2 \big( \gamma_{+}(t) \abs{U^{+}_2(t)}^2  - \abs{\gamma_{-}(t)} \abs{U^{-}_2(t)}^2 \big).
\end{equation}
Notice that for this result no numerical computation was involved, so we are not concerned with a specific choice of the size of the Hilbert space.
We show the results in Fig.~\ref{fig: p_divisibility} for the triplet state $\ket{\Psi_+}$, where we can see that one eigenvalues becomes negative for specific time windows which depend on the parameters chosen, and in Fig.~\ref{fig: p_div_new}

\begin{figure}
    \centering
    \includegraphics[width=0.99\linewidth]{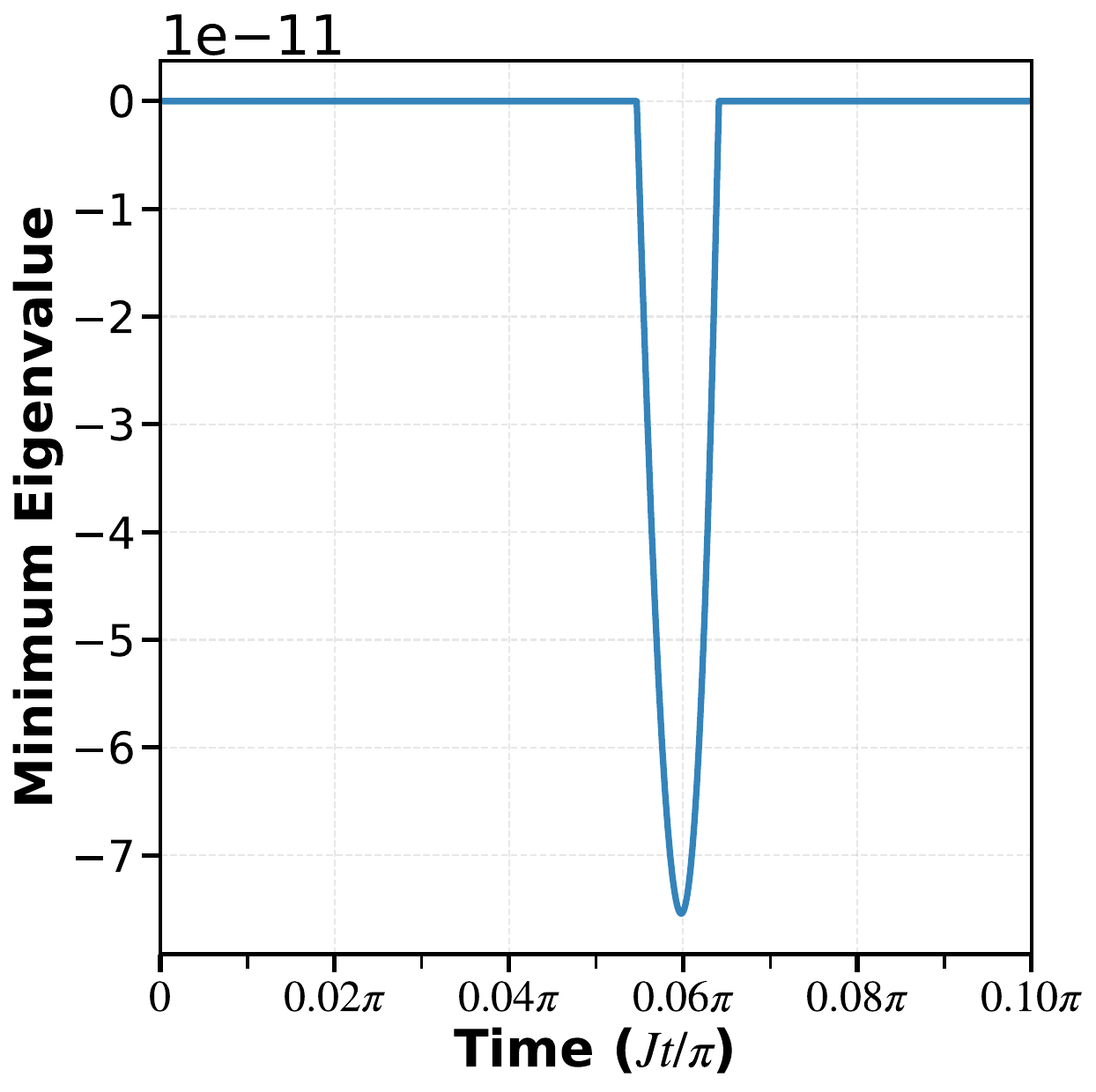}
    \caption{Time evolution of the minimum eigenvalue of the density matrix, computed numerically from the master equation in Eq.~(\ref{eq: non_diag_canonical_form}), corresponding to the initial states with distinct spin. The parameters are the same as in Fig.~\ref{fig: eigenvalues}, corresponding to the on-resonance case with $\omega_0 = J$. Since both the time interval in which the eigenvalue is negative and its absolute value are small, we can assume that the derived master equation is physical for the parameter choice of the main text.}
    \label{fig: density_matrix_eigenvalues}
\end{figure}

We now move to investigate the case for $\ket{\Phi}$. The calculations are more cumbersome, and the full expression of the matrix $W_J^{\ket{\phi}}(t)$ of Eq.~(\ref{WJmatrix}) is
\begin{equation}
    \begin{split}
        & W_J^{\ket{\Phi}}(t)= \gamma_0 (t) (\ket{00}\bra{00} +\ket{11}\bra{11} +\ket{10}\bra{10} +\ket{01}\bra{01}) \\
        & \sum_{\pm} \gamma_{\pm}(t) \bigg[ \abs{U^{\pm}_1(t) + i U^{\pm}_2(t)}^2 (\ket{01}\bra{10} + \ket{10}\bra{01} \\
        & +\ket{01}\bra{01} + \ket{10}\bra{10})\\
        & - \abs{U^{\pm}_2(t)}^2 (\ket{02}\bra{02} + \ket{20}\bra{20} \\
        & - \ket{02}\bra{20} - \ket{20}\bra{02})\\
        & + (U^{\pm}_1(t) + i U^{\pm}_2(t))(- i U^{\pm *}_2(t))(\ket{01}\bra{02} - \ket{01}\bra{20} \\
        & + \ket{10}\bra{02} - \ket{10}\bra{20})\\
        & + (U^{\pm}_1(t) - i U^{\pm *}_2(t))(+ i U^{\pm}_2(t))(\ket{02}\bra{01} - \ket{20}\bra{01} \\
        &+ \ket{02}\bra{10} - \ket{20}\bra{10})\bigg].
    \end{split}
\end{equation}
This matrix can be diagonalized numerically, and we can retrieve its eigenvalues to check for violations of p-divisibility. We build a Hilbert space of size $d=9=3^2$, where each spatial regions has size $3$. 
Moving forward, when deriving an approximated evolution map for a reduced system, it is crucial to determine whether the density matrix remains physical at all times. We already know that the map is trace-preserving and Hermiticity-preserving. However, we still need to check whether the density matrix remains positive definite also in the non-Markovian regions where P-divisibility is violated. To do this, we can numerically compute the eigenvalues of the density matrix at all times, for the parameters choice adopted in the main text, and check whether they are positive or not. In Fig.~\ref{fig: density_matrix_eigenvalues}, we show the time evolution of the smallest eigenvalue of the density matrix for a specific set of parameters, corresponding to the off-resonance scenario. We can see that the minimum eigenvalue is slightly negative only for a very short time window at the beginning of the evolution, while it remains positive for the rest of the dynamics inside numerical. This confirms that the approximated map we derived is mostly physical for the initial states that we considered, despite the non-Markovian nature of the dynamics. 

\section{Numerically comparing the master equation with the unitary closed system evolution}
\label{sec: pseudo_mode}
In this section, we want to check how well the master equation we have shown in Eq.~(\ref{eq: canonical_form}) compares with the exact reduced dynamics that is obtained via the partial trace applied to the unitary evolution of the total system. 
However, since an analytical derivation is not feasible, another option is to numerically compute the evolution of the total density matrix. This might prove extremely hard, since we have assumed that the system is coupled to a continuum of frequencies. One alternative to this is the pseudomode method. The key idea is that a reservoir with a Lorentzian spectral density can be mapped onto a single effective harmonic mode (the pseudomode) which itself decays into a memoryless Markovian bath. The system then interacts only with this pseudomode(s), and the full non-Markovian dynamics of the system is recovered by tracing them out. For more details, we refer the interested reader to the wide and detailed literature on this subject \cite{garraway1997nonperturbative, dalton2001theory, mazzola2009pseudomodes, tamascelli2018nonperturbative, chen2019markovian, mascherpa2020optimized, pleasance2020generalized, menczel2024non, zhou2024systematic}. In our case, our system is coupled to two local environments with Lorentzian spectral density, thus we require two effective bosonic pseudomodes. We label their respective creation and annihilation operators as $\hat{c}^{\dagger}_{X, pm}$ and $\hat{c}_{X, pm}$, with $X = \{ L, R\}$. More specifically, the coupling occurs through the local dephasing operator $\sigma_{z, X}$. We can write the system-pseudomode Hamiltonian as
\begin{equation}
\hat{H}_{pm} = \hat{H}_S + \hat{H}_D + \sum_{X = \{ L, R\}} \left[ \omega_{pm} \hat{c}^{\dagger}_X \hat{c}_X + g_{pm} \sigma_{z, X} (\hat{c}^{\dagger}_X + \hat{c}_X)\right] \text{.}
\end{equation}
The two pseudomodes are damped with rate $\gamma_{pm}$ into two flat background reservoirs, so that the total master equation in Schr\"{o}dinger picture becomes
\begin{equation}
    \dot{\hat{\varrho}}_{S+pm}^T = -i[H_{pm}, \hat{\varrho}_{S+pm}^T] + \sum_{X = \{ L, R\}} \gamma_{pm} D[\sigma_{z,X}].
\end{equation}
The three parameters per pseudomode, $\gamma_{pm}, \omega_{pm}, g_{pm}$, are fixed by matching the pseudomodes' Green's function with the original bath correlation function: if the two coincide, the reduced dynamics obtained from $\hat{\varrho}_{S+pm}$ will coincide with the exact reduced state of the original unitary configuration. The pseudomodes' Green's function is
\begin{equation}
    C_{pm} (t) = g_{pm}^2 e^{- (\gamma_{pm} + i \omega_{pm})t},
\end{equation}
while the original correlation function is the one we have derived in the previous section,
\begin{equation}
     \langle \el(t) \el(s) \rangle = g_0 e^{-(i \omega_0 + \lambda) \abs{t - s}},  \;\;\; (t \geq s),
\end{equation}
so that we have
\begin{equation}
    \begin{split}
        & \omega_{pm} = \omega_0, \\
        & \gamma_{pm} = \lambda, \\
        & g_{pm} = \sqrt{g_0}.
    \end{split}
\end{equation}

\begin{figure}
    \centering
    \includegraphics[width=0.99\linewidth]{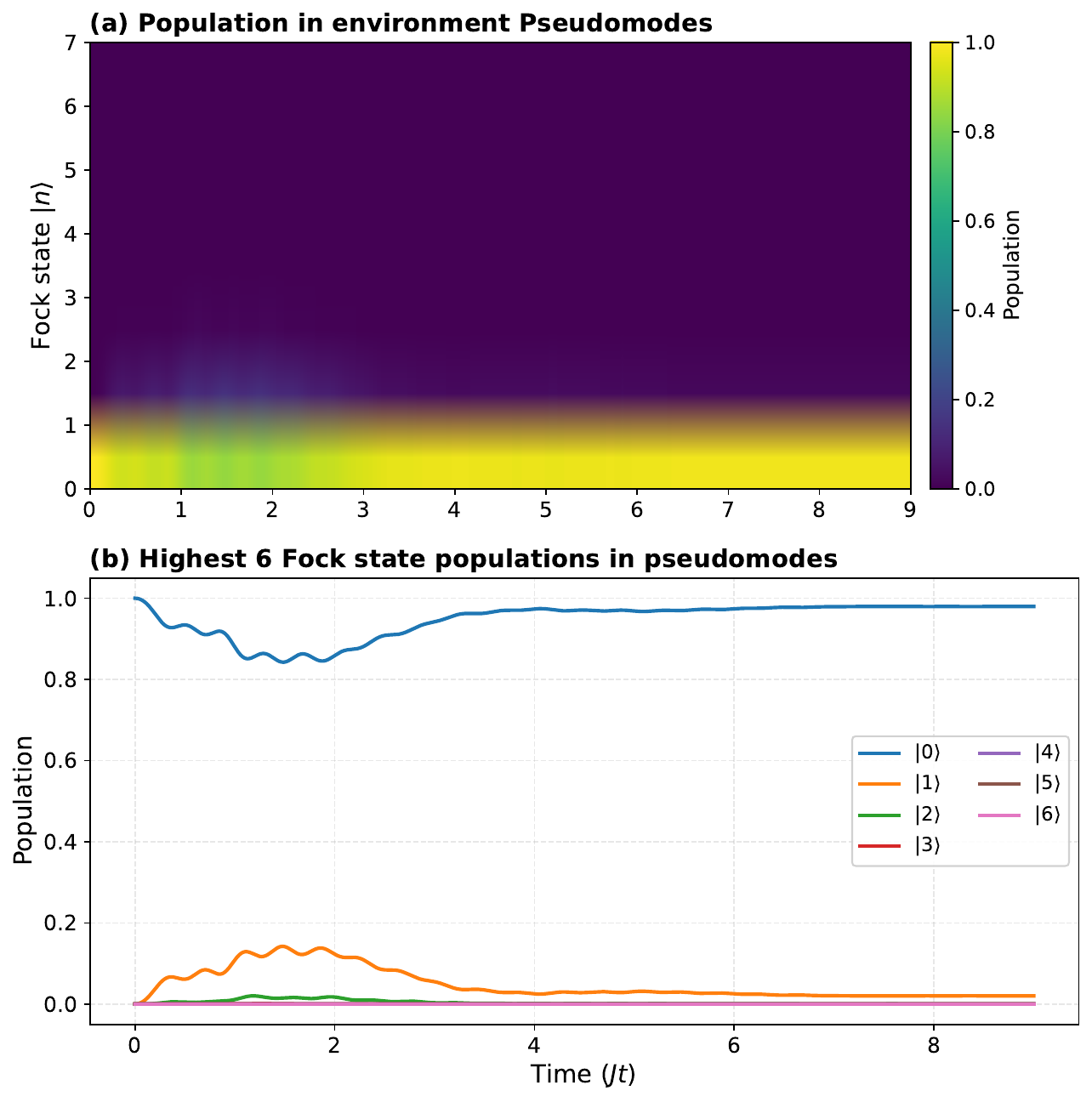}
    \caption{Example of time evolution of the populations of the two pseudomodes, for a cutoff value equal to $7$. $\lambda = 0.5, g_0 = 0.1, J = 2.5, \omega_0 = 2.5$. The steady state population of the highest level $\ket{6}$ is $P_{\ket{6}} \approx 9\cdot10^{-12}$}
    \label{fig: cutoff}
\end{figure}

In our framework, this construction has several advantages: (i) it provides an exact embedding for Lorentzian environments (and, more generally, for sums of Lorentzians by introducing multiple pseudomodes), (ii) it yields a time-local Lindblad equation in the enlarged space, which is numerically efficient to simulate, and (iii) it offers a transparent physical picture of non-Markovianity as arising from coherent back-action mediated by a small number of effective modes. As a result, we have a reliable numerical benchmark for our newly derived master equation. To ensure that the two approaches are comparable, we need to trace out the pseudomodes of the total density matrix $\hat{\varrho}_{S+pm}^T$ at each time step, so that we can compare it with the reduced density matrix $\hat{\varrho}_S(t)$ obtained through the master equation. Additionally, we need to make sure that the numerical dimension of the Hilbert space for the pseudomodes is high enough to avoid any spurious effects.
In Fig.~\ref{fig: cutoff}, we show the populations of the pseudomodes over time, and we see that the highest allowed levels are not significantly populated, which justifies the truncation of the energy level at $N=8$ that we used in our investigation.

\section{Dynamical evolution of superposition states}
\label{sec: example}
In Sec.~\ref{sec: results}, we have studied the dynamical evolution of two specific two-particle initial states with the aim of investigating the consequences of dephasing noise in the Hong-Ou-Mandel effect and in the entanglement generation protocol. Here, we report two extra examples of dynamical evolution. First, we consider the following initial state
\begin{alignb}
    \ket{\chi (0)}& = \frac{1}{2}(\ket{1}_L + \ket{2}_L) \otimes (\ket{1}_R + \ket{2}_R) \\
    & = \frac{1}{2}(\ket{1}_L\ket{1}_R + \ket{1}_L\ket{2}_R + \ket{2}_L\ket{1}_R + \ket{2}_L\ket{2}_R)
\end{alignb}
where, as we did in the previous section for equal spin bosons, we define as $\ket{n}_X$ the state vector representing $n$ equal-spin particles in the $X$ spatial region.
This corresponds to a tensor product of two pure states, initially in a \textit{local} superposition of bosonic Fock states with one and two particles with identical spin. This will support validating the derived master equation when the initial state, during its evolution, explores sectors of the Hilbert space with different total numbers of particles.

We look at the evolution of the first order correlation $\mathcal{G}_1(t) = |\langle \lsig \hat{R}_{\sigma}^{\dagger} + \text{h.c.}\rangle|(t)$. This observable commutes with the tunneling dynamics, which implies that under noiseless dynamics its average value is constant (equal to one, for this specific initial state).

\begin{figure}
    \centering
    \includegraphics[width=1\linewidth]{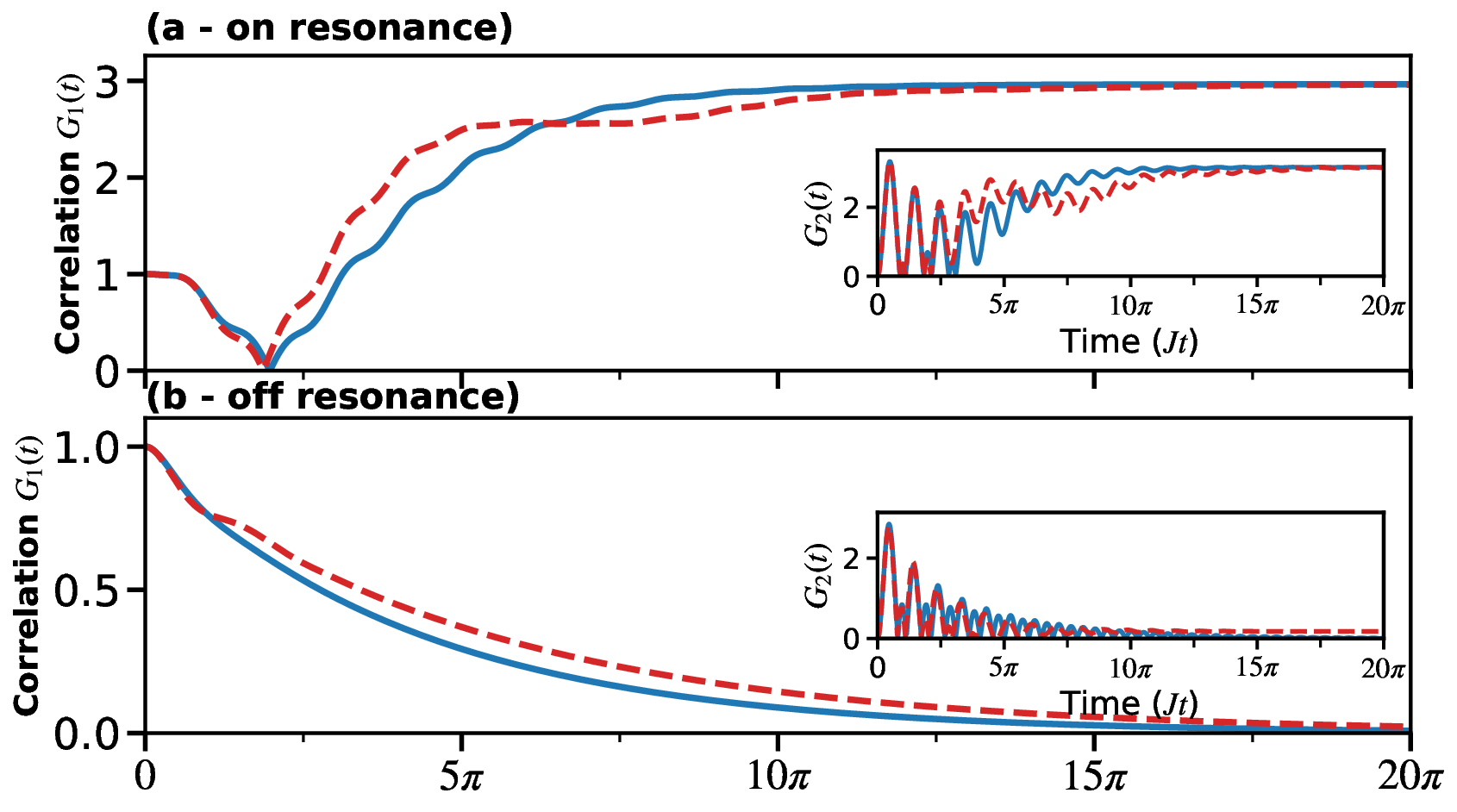}
    \caption{On-resonance (a) and off-resonance (b) evolution of first-order correlation function for equal-pseudospin bosons from numerical simulations using the full master equation (solid blue line) and the unitary pseudomode evolution (dashed red line). The parameters correspond to those used in the main text, as in Fig.~\ref{fig: OFF_spinless} and in Fig.~\ref{fig: ON_RES_spinless}. The intial state is $\ket{\chi (0)} = \frac{1}{2}(\ket{1}_L + \ket{2}_L) \otimes (\ket{1}_R + \ket{2}_R)$}
    \label{fig: multi-particle}
\end{figure}

The qualitative dynamical difference between the two distinct regimes can be clearly seen in Fig.~\ref{fig: multi-particle}. Off-resonance, and in line with expectations of white noise phase diffusion, the coherence between the two spatial regions exponentially decays to zero. On-resonance, instead, the coherent oscillatory dynamics, stabilized by resonant dephasing, brings to a revival of quantum coherence, followed by a stabilization into a correlated steady state.

\bibliography{ref.bib}

\newpage

\end{document}